\documentclass[aps,10pt,pre,floatfix,twocolumn,superscriptaddress]{revtex4-2}

\usepackage{color}
\usepackage{graphicx}
\usepackage{subfigure,amsmath, amsthm, amssymb}
\usepackage[colorlinks,linkcolor=blue,citecolor=blue]{hyperref}
\newcommand{\be}{\begin{equation}}
\newcommand{\ee}{\end{equation}}
\newcommand{\bea}{\begin{eqnarray}}
\newcommand{\eea}{\end{eqnarray}}

\begin{document}
\title{Lattice models for ballistic aggregation: cluster-shape dependent exponents}
\author{Fahad Puthalath}
\email{fahad.puthalath@dlr.de}
\affiliation{Institut für Materialphysik im Weltraum, Deutsches Zentrum für Luft- und Raumfahrt (DLR), 51170 Köln, Germany}
\affiliation{Institut für Theoretische Physik, Universität zu K\"{o}ln, Zülpicher Strasse 77, 50937 Köln, Germany}

\author{Apurba Biswas}
\email{apurbab@imsc.res.in}
\affiliation{The Institute of Mathematical Sciences, C.I.T. Campus, Taramani, Chennai 600113, India}
\affiliation{Homi Bhabha National Institute, Training School Complex, Anushakti Nagar, Mumbai 400094, India}

\author{V. V. Prasad}
\email{{prasadvv@cusat.ac.in}}
\affiliation{{Department of Physics, Cochin University of Science and Technology, Cochin -
682022 India}}

\author{R. Rajesh}
\email{rrajesh@imsc.res.in}
\affiliation{The Institute of Mathematical Sciences, C.I.T. Campus, Taramani, Chennai 600113, India}
\affiliation{Homi Bhabha National Institute, Training School Complex, Anushakti Nagar, Mumbai 400094, India}
  
\date{\today}

\begin{abstract}

We study ballistic aggregation on a two dimensional square lattice, where particles move ballistically in between momentum and mass conserving coalescing collisions. Three models are studied based on the shapes of the aggregates: in the first the aggregates remain point particles, in the second they retain the fractal shape at the time of collision, and in the third they assume a spherical shape. The exponents describing the power law  temporal decay of number of particles and energy as well as dependence of velocity correlations on mass are determined using large scale Monte Carlo simulations. It is shown that the exponents are universal only for the point particle model. In the other two cases, the exponents are dependent on the initial number density and correlations vanish at high number densities. The fractal dimension for the second model  is close to 1.49.
\end{abstract}

\maketitle

\section{Introduction}

There is a wide variety of physical phenomena at different length scales in which aggregation of particles/clusters to form larger particles is the predominant dynamical process~\cite{whitesides2002self}. Examples include aerosols~\cite{hidy1970dynamics,drake1972topics}, agglomeration of soot ~\cite{friedlander2000smoke,sorensen2018light}, gelation~\cite{stockmayer1943theory}, cloud formation~\cite{cloud-1997}, colloidal agrregates~\cite{ganesh2018colloidal}, astrophysical problems~\cite{lee-astro-2000}, aggregation of dust particles in planetary discs~\cite{esposito2006planetary,brilliantov2015size,blum2018dust}, dynamics of Saturn's rings~\cite{brilliantov2015size,connaughton2018stationary}, polyelectrolytes~\cite{tom2016aggregation,tom2017aggregation}, networks~\cite{dorogovtsev2002evolution}, etc.  A minimal model that focuses only on the effects of aggregation is the cluster-cluster aggregation (CCA) model in which particles that come into contact undergo mass conserving coalescence (reviews may be found in Refs.~\cite{leyvraz2003scaling,handbook,aldous1999deterministic}). In addition to its relevance for different physical phenomena, CCA has also been studied as a nonequilibrium system undergoing scale invariant dynamics that is described by exponents that depend only on very generic features of the transport process. This universal feature allows applications of results for CCA in seemingly unrelated systems like Burgers turbulence~\cite{kida1979asymptotic,frachebourg2000ballistic,tribe2000large,burgers2013nonlinear,dey2011lattice}, Kolmogorov self-similar scaling~\cite{takayasu1989steady,connaughton2005breakdown,connaughton2006cluster}, granular systems~\cite{ben-naim-shock-like-1999,nie2002dynamics,rajesh-granular-prl-2014,shinde2009equivalence}, hydrodynamics of  run and tumble particles~\cite{dandekar2020hard}, evolution of planetesimals~\cite{wetherill1988formation}, geophysical flows~\cite{meakin1991fractal}, etc.

Among the different transport processes, ballistic transport is of particular importance and the resultant CCA is known as the ballistic aggregation (BA) model, the focus of this paper.  In the BA model,  momentum is additionally conserved in collisions. The BA model with spherical particles has been studied using mean field theory, large scale simulations in two and three dimensions, and is exactly solvable in one dimension. It is found that the number of particles, $n(t)$, and energy, $e(t)$, decrease with time, $t$, as power-laws:  $n(t)\propto t^{-\theta_n}$, $e(t)\propto t^{-\theta_e}$. These exponents have been determined in $d$-dimensions within a mean field approximation which assumes that the particle density is small,  that the particles are compact spherical clusters of equal density, and that the velocities of the particles constituting a cluster are uncorrelated. Within these assumptions, scaling arguments predict the existence of a growing length
scale ${\mathcal L}_t \sim t^{1/z^{\mathrm{mf}}}$ with 
$z^{\mathrm{mf}}= (d+2)/2d$ and mean field exponents, 
$\theta^{\mathrm{mf}}_n=2d/(d+2)$ and  $\theta^{\mathrm{mf}}_e=\theta^{\mathrm{mf}}_n$~\cite{Carnevale1990}. The correlations in the initial velocities of the constituents of a cluster is characterized by an exponent $\eta$: $\langle v_m^2 \rangle \sim m^{-\eta}$, where $\langle v_m^2 \rangle$ is the mean square velocity of a particle of mass $m$. In the mean field approximation, by assumption, $\eta^{\mathrm{mf}}=1$. The mean field results for the exponents are of particular significance to the study of the unrelated problem of freely cooling granular gas in which ballistic particles undergo energy-dissipating, momentum conserving binary collisions. It has been shown that exponent characterizing the  energy decay in the granular gas is equal to $\theta_e^{\mathrm{mf}}$ in dimensions upto three~\cite{ben-naim-shock-like-1999,rajesh-granular-prl-2014,pathak2014inhomogeneous}.

In one dimension, BA is exactly solvable and the exponents match with the mean-field exponents~\cite{piasecki1992universal,frachebourg1999exact,frachebourg2000ballistic,ben2000stochastic}. However, in two and three dimensions, it has been shown that the exponents for BA with spherical particles depend on the initial number density $n_0$. In two dimensions and for dilute systems ($n_0 \to 0$), it has been shown that the numerically  obtained $\theta_n$ is $17$\% larger than 
$\theta_{n}^{\mathrm{mf}}$  because of 
strong velocity correlations between colliding 
aggregates, with $\eta$ decreasing from $\eta \approx 1.33$ for low densities to $\eta \approx 1=\eta^{\mathrm{mf}}$ for high densities~\cite{rajesh-granular-prl-2014,Trizac-PRL-1995,Trizac1996,Trizac2003,Subhajit2018}. In three dimensions, it is found that  as $n_0$ increases from $0.005$ to $0.208$, $\theta_e$ decreases from $\theta_e=1.283$ to
$1.206$ and appears to converge to the $\theta_e^{\mathrm{mf}}=1.2$
with increasing $n_0$, and $\eta$ decreases from $\eta \approx 1.23$ for low densities to $\eta \approx 1=\eta^{\mathrm{mf}}$ for high densities~\cite{rajesh-granular-prl-2014,Subhajit2018}. It is remarkable that the mean field results describe well only the systems with large $n_0$, while its derivation assumes the limit $n_0 \to 0$. This counterintuitive result has been argued to be due to the randomization of the velocities at higher densities due to avalanche of coagulation events that occur due to the overlap of  a newly created spherical particle with already existing particles, as the number density is increased.

While the kinetics of BA with spherical particles are reasonably understood, much less is known for the exponents when clusters have non-spherical shapes. The  scaling analysis can be extended to the case when the mass scales with radius with a fractal dimension $d_f$~\cite{Trizac1996} (also see Sec.~\ref{sec:scaling} where we review scaling theory). The scaling theory leads to hyperscaling relations between the different exponents independent of the mean field assumptions. Fractal shapes are of particular importance in the case of the experiments on aggregates of soot~\cite{sorensen2018light,zhang2020three}, mammary epithelial cells~\cite{leggett2019motility,liu2021scale}, spray flames~\cite{simmler2022characterization}, etc., where the aggregates have a fractal dimension different from those of compact structures ($d_f= 2, 3$). While the fractal dimensions seen in experiments~\cite{sorensen2018light,leggett2019motility} are sometimes close to that for diffusion limited aggregation ($d_f \approx 1.7$), there are many examples for which it is very different (for instance $1.54$ for sprays~\cite{simmler2022characterization},  $1.5$ for cells~\cite{liu2021scale} or $2.4$ for soot~\cite{,zhang2020three}). The fractal dimension of aggregates formed by ballistic motion is not known to the best of our knowledge. In addition, it is also not known how the exponents for BA change when the shape of the clusters deviates from spherical. Neither is it known whether the mean field limit is reached for any particular limit of number density when the clusters are fractal. Finally, in the characterization of mass distribution, a relevant exponent is the scaling of mass distribution with small mass, namely $N(m) \sim m^{\zeta}$ [also see definition in Eq.~(\ref{eqn:small_mass_Nmt})]. The exponent $\zeta$ is an independent exponent and cannot be obtained from scaling theory, and is not known even for BA with spherical particles.

To answer these questions, we study three differently shaped clusters (named as models A, B, C) undergoing BA on the square lattice. We choose a lattice approach as it allows us to maintain fractal shapes in a computationally efficient manner. Lattice models are known to reproduce the same results as the continuum for BA in one dimension~\cite{ostojic2004clustering,dey2011lattice}, and we expect the equivalence to hold true for two and higher dimensions.  In model A, the clusters occupy a single site irrespective of its mass. This limiting model allows us to separate the dependence of the velocity correlations on the initial density from the dependence on mass-dependent shape. In model B, we study clusters where the clusters maintain the shape at the time of contact.  Such clusters turn out to be fractal. In model C, we study ``spherical" clusters in which the lattice approximation to the disc is maintained. This model allows us to study lattice effects by comparing the results on the lattice with the continuum results. In addition, we obtain the value of the exponent $\zeta$ for all the three models. The results for the three models are summarized in Table~\ref{table_mod1} (model A), Table~\ref{table_mod2} and Fig.~\ref{fig:mod2_exp_dendep} (model B), Table~\ref{table_mod3} and Fig.~\ref{fig:mod3_exp_dendep} (model C). For model A, we show that the exponents are universal, in the sense that it is independent of the initial number density, $n_0$ and it is different from the mean field results. For models B and C, we find that the exponents are dependent on $n_0$ and approach the mean field assumptions of uncorrelated velocities only in the limit of large $n_0$. The fractal dimension for model B, on the other hand, is universal, with $d_f \approx 1.49$.

The remainder of the paper is organized as follows. Section~\ref{model} contains a definition of the different models as well as a description of the  simulation methods. We briefly review the scaling theory for BA with differently shaped particles in Sec.~\ref{sec:scaling}. In Sec.~\ref{sec:results}, for the three models, we describe the  results for the different exponents obtained from large scale Monte Carlo simulations. Section~\ref{sec:conclusion} contains a summary and discussion of the results.

\section{Model \label{model}} 

In this section, we define the three models that we study in this paper. Consider a square lattice of size $L\times L$ with periodic boundary conditions. Initially $N$ particles, each of mass $1$, are randomly distributed  with a site having utmost one particle. Each particle is assigned a velocity whose magnitude is drawn from a uniform distribution in $[0,1)$ and whose direction is chosen uniformly in $[0, 2 \pi)$.  The velocity of the center of mass  is set to be zero by choosing an appropriate frame of reference.  
The system evolves stochastically in time as follows. A particle with velocity $(v_x, v_y)$ hops in the $x$-direction with rate $|v_x|$ in the positive (negative) direction depending on whether $v_x$ is positive (negative). Likewise, it hops along the $y$-axis with rate 
$|v_y|$ in the direction determined by the sign of $v_y$. When two particles collide, they aggregate to form a new particle. The mass of the new particle is the sum of the constituent particles while the new velocity is determined by conservation of linear momentum.  The shape of the new particle is determined based on three different rules, leading to three different models.

\subsection{Model A: Point particles}

In model $A$, when a particle hops onto a site which is already occupied, then the two particles coalesce, conserving mass and momentum. The new particle occupies the same lattice site. We call this model the point particle model, since the sizes of all the particles are the same (one lattice site) irrespective of their mass. The model is motivated from its similar counterpart in one dimension as considered in Ref.~\cite{dey2011lattice}.

\subsection{Model B: Fractal clusters}

\begin{figure}
 \includegraphics[width=\columnwidth]{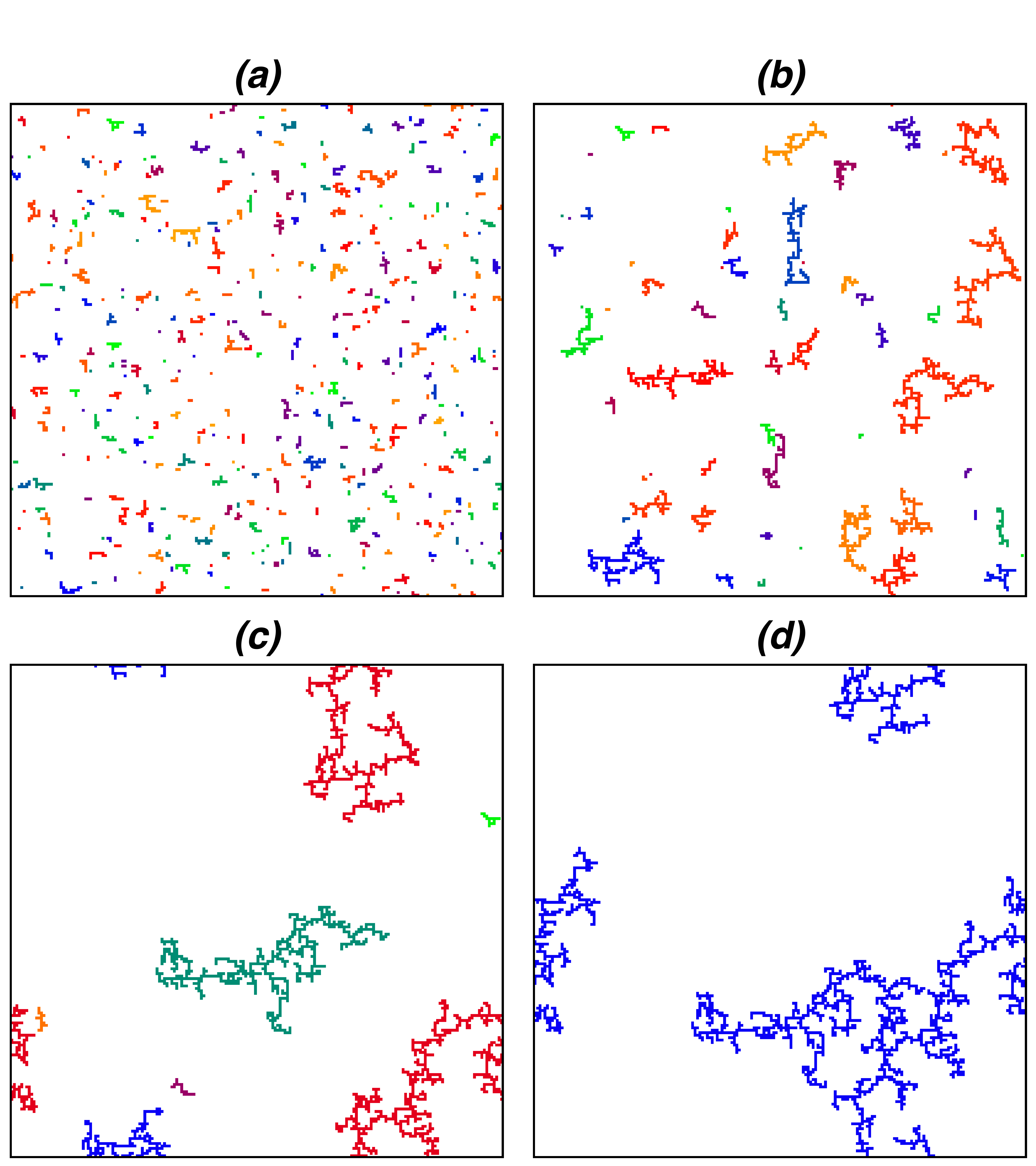}
 \caption{\label{fig:modelB} Snapshots of the  configurations at different times $t$  for model B (fractal clusters), where number of clusters decrease with time. The different panels correspond to (a) $t=50$, (b) $t=500$, (c) $t=5000$ and (d) $t=25790$. The data are for system size $L=200$ and initial number of  $N=2000$ particles ($n_0=0.05$).}
\end{figure}

In model $B$, the particles, also referred to as clusters,  are extended objects consisting of a collection of sites that are linked to each other by nearest neighbor bonds. When a cluster hops,  if any of the lattice sites belonging to it becomes adjacent to a site belonging to another cluster, then the two clusters coalesce.  The new cluster maintains the shape at the time of coalescing, till it collides with another cluster at a future time.  The new velocity of the cluster is determined through momentum conservation.  Snapshots of the configuration at different times are shown in Fig.~\ref{fig:modelB}. The clusters are extended and will be shown to be fractals. The model B is motivated from the study of aggregation in two dimensions in the continuum model as considered in Ref.~\cite{PhysRevLett.51.1119}.

\subsection{Model C: Spherical clusters}

\begin{figure}
 \includegraphics[width=\columnwidth]{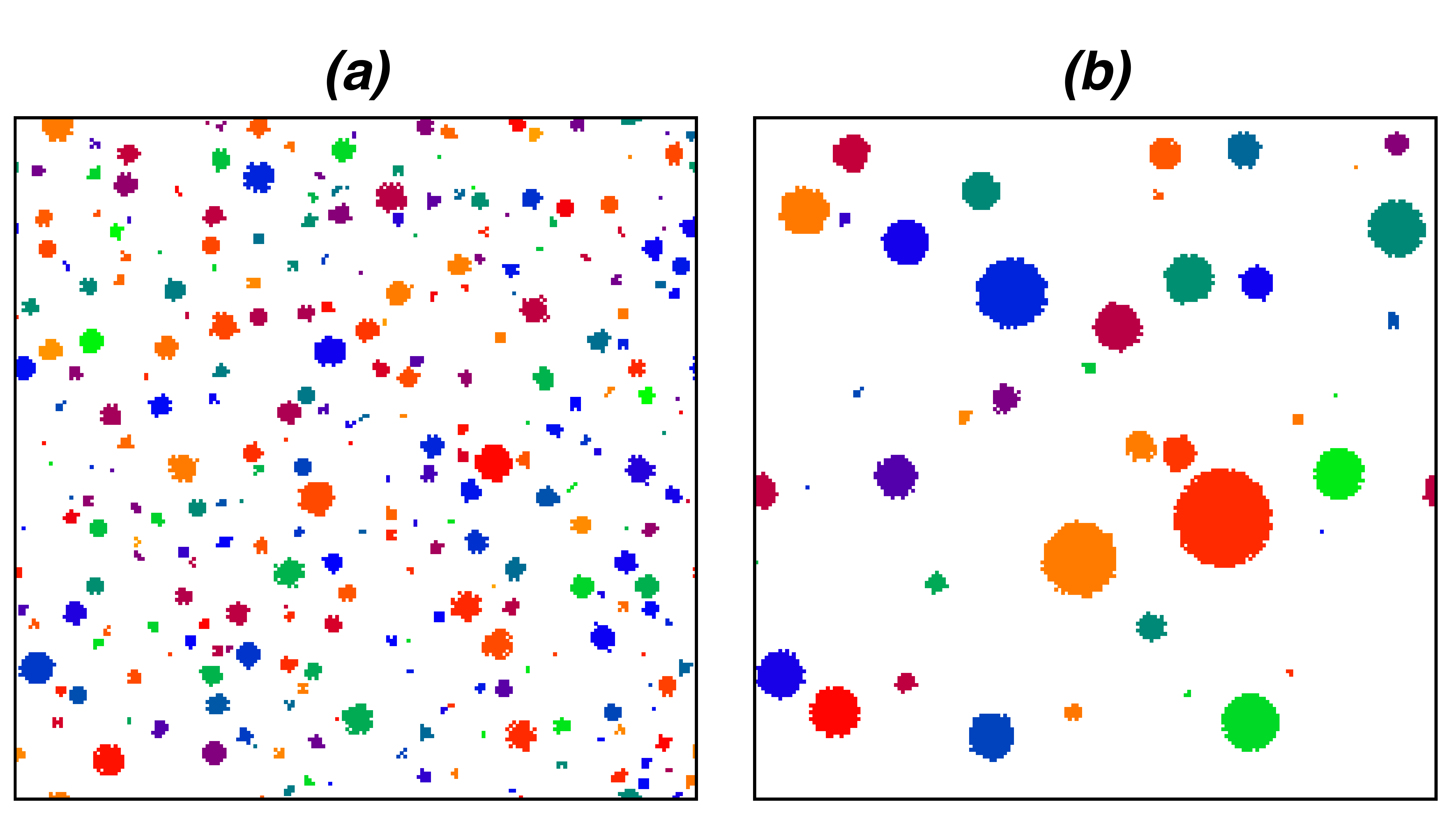}
 \caption{\label{fig:modelC}Snapshots of the  configurations at different times $t$  for model C (spherical clusters), where number of clusters decrease with time. The different panels correspond to (a) $t=100$ and (b) $t=500$. The data are for system size $L=200$ and initial number of  $N=4000$ particles ($n_0=0.1$).}
\end{figure}

In model $C$, like in model $B$, particles are extended clusters. However, the shape of these particles are constrained to be spherical. When two particles come into contact, they are replaced by a new spherical particle. The center of mass of the new particle is chosen to be lattice site closest to  the center of mass of the constituent particles. To construct a spherical cluster on the square lattice, we fill all lattice sites within circles of increasing radius. The sites in the outermost shell, if not fully occupied, are chosen  at random. This rearrangement of sites to form a spherical shape will, at times, lead to the new cluster overlapping with other nearby clusters, triggering an avalanche of coalescence events. The model C is motivated from its similar counterpart in the continuum model as considered in Ref.~\cite{rajesh-granular-prl-2014,Trizac-PRL-1995,Trizac1996,Trizac2003,Subhajit2018}. Snapshots of a typical time evolution are shown in Fig.~\ref{fig:modelC}.

\subsection{Details of simulation}

The models are simulated using standard Monte Carlo methods. In all the models, the rates of hopping are modified when mass of particles change during aggregation. In addition, the number of particles decrease with aggregation. To make the simulations efficient, we use a variable time step that changes according to the total rate of hopping of all the particles. In particular, we choose a time step such that the probability of one of the particles hopping is exactly equal to $1$. This makes the simulations rejection-free.

In model B and model C, where extended clusters hop as a single unit, we identify the different clusters and their merging using the well-known Hoshen-Kopelman algorithm~\cite{hk1976}. Simulations were carried out for different system sizes varying from $L = 100$ upto $L = 10000$ for all three models and a wide range of number densities $n_0$. The simulation is continued till all the clusters aggregate together to form the final single cluster. The details of the densities and the lattice sizes used for simulations of the three models are given in Table~\ref{table_sim_details}.
\begin{table}
\caption{\label{table_sim_details}Simulation details.}
		\begin{ruledtabular}
		\begin{tabular}{c l l}
		Model & $L$'s simulated & Number densities ($n_0$)\\
		\hline
		A & upto 1000 & 0.01 - 1.00\\
		B & upto 10000 & 0.001 - 0.01\\
		C & upto 10000 & 0.0001 - 0.16\\
	\end{tabular}
	\end{ruledtabular}
\end{table}

\section{\label{sec:scaling} Review of scaling theory}

In this section, we review the scaling theory for BA, described initially in
Ref.~\cite{Carnevale1990}.  Here, we give a scaling argument based on the  Smoluchowski equation for aggregation (see Refs.~\cite{leyvraz2003scaling,handbook} for reviews).  Different scaling arguments, leading to the same results, may be found in Refs.~\cite{Trizac1996,Trizac2003}. Let $N(m,t)$ denote the average density of clusters of mass $m$ at time $t$. $N(m,t)$ evolves in time as
\begin{align}
&\frac{d N(m,t)}{d t} = -N(m,t) \int_{0}^{\infty}dm_1K(m,m_1)N(m_1,t) \nonumber \\
&+ \frac{1}{2}\int_{0}^{m}dmK(m_1,m-m_1)N(m,t)N(m-m_1,t), \label{eqn:finalsme} 
\end{align}
where the kernel $K(m_1,m_2)$ is the rate at which particles of masses $m_1$ and $m_2$ collide. The first term in the right hand side of Eq.~(\ref{eqn:finalsme}) describes a loss term where a particle of mass $m$ collides with another particle, while the second term describes a gain term where two particles collide  to form a particle of mass $m$.

We restrict ourselves to homogeneous kernels, which are known to describe many physical systems, examples of which may be found in Refs.~\cite{leyvraz2003scaling,handbook}. Homogeneous kernels have the property
\be
K(h m_1, h m_2)=h^\lambda K(m_1,m_2), ~~h>0,
\label{eqn:kernelhmg}
\ee
where $\lambda$ is called the homogeneity exponent. 
For $\lambda<1$, and for large masses and times, it can be shown that Eq.~(\ref{eqn:finalsme}) is solved by a $N(m,t)$ which has the scaling form
\be
N(m,t) \simeq \frac{1}{t^{2\theta_n}}\Phi\bigg(\frac{m}{t^{\theta_n}}\bigg).
\label{eqn:N(m,t)_final}
\ee
For $x\gg1$, $\Phi(x)$ vanishes exponentially. For $x\ll1$, $\Phi(x)$ is a power law 
\be
\Phi(x) \sim x^{\zeta},~~x\ll 1.\label{eqn:zeta_form}
\ee
Thus, there are two exponents $\theta_n$ and $\zeta$ characterizing the mass distribution $N(m,t)$.  

The exponent $\theta_n$ describes how the mean density of particles $n(t) = \int_m N(m,t) dm$ decreases with time. Integrating Eq.~(\ref{eqn:N(m,t)_final}), we obtain
\be 
n(t) \sim t^{-\theta_n}.
\ee
The exponent $\zeta$ describes the power law dependence of $N(m,t)$ on mass for small masses:
\be
N(m,t) \sim \frac{m^\zeta}{t^{\theta_n(2+\zeta)}},~~ m \ll t^{\theta_n}.\label{eqn:small_mass_Nmt}
\ee

The dependence of $\theta_n$ on the homogeneity exponent $\lambda$ can be obtained by substituting Eq.~(\ref{eqn:N(m,t)_final}) into Eq.~(\ref{eqn:finalsme}), and is known to be (for example, see Refs.~\cite{leyvraz2003scaling,handbook})
\be 
\theta_n = \frac{1}{1-\lambda}.
\label{eqn:theta_of_lambda}
\ee

We now focus on the collision kernel that corresponds to BA. Assuming a homogeneous mixture of clusters of all masses, the rate of collision between two masses $m_1$ and $m_2$  is  proportional to  $(r_1+r_2)^{d-1}|\vec{v_1}-\vec{v_2}|$ where $r_1$ and $r_2$ are the radii  of the particles,
$\vec{v_1}$ and  $\vec{v_2}$ the velocities, and $d$ is the dimension. The relative velocity may be approximated as
$|\vec{v_1}-\vec{v_2}|\approx \sqrt{v_1^2+v_2^2}$. Thus, the collision kernel for BA may be written as
\be 
K(m_1,m_2) \propto (r_1+r_2)^{d-1} \sqrt{v_1^2+v_2^2}. \label{eqn:kernel1}
\ee
To express the radii and velocities in terms of the masses, we assume that the typical speed, $v_m$, of particles of mass $m$,  scales with mass as 
\be
v_m^2\sim m^{-\eta}.
\label{eq:eta-defn}
\ee
The radii are related to mass though the fractal dimension, $d_f$, of a cluster:
\be
r \propto m^{1/d_f}.
\ee
Thus, the kernel in Eq.~(\ref{eqn:kernel1}) reduces to
\be 
K(m_1,m_2) \propto \left[m_1^{1/d_f}+m_2^{1/d_f}\right]^{d-1} \sqrt{m_1^{-\eta}+m_2^{-\eta}}.
 \label{eqn:our_kernel}
\ee

This kernel is homogeneous in its arguments with homogeneity exponent given by
\be 
\lambda=\frac{d-1}{d_f}-\frac{\eta}{2}.\label{eqn:lambda}
\ee
From Eq.~(\ref{eqn:theta_of_lambda}), we then  obtain
\be 
\theta_n = \frac{2d_f}{2d_f-2(d-1)+\eta d_f}.
\label{eqn:thea_general}
\ee

Another quantity of interest is the mean kinetic energy $e(t)$, defined as
\be
e(t)  \simeq \int dm \frac{1}{2} m v_m^2 N(m,t).
\ee
The energy density decreases in time as a power law $e(t) \sim t^{-\theta_e}$. Substituting $v_m^2 \sim m^{-\eta}$, we obtain the scaling relation
\be
\theta_e = \eta \theta_n.
\label{eqn:thetabeta_foralpha}
\ee

We now reproduce the results obtained for BA in  Ref.~\cite{Carnevale1990} which we refer to as the mean field BA exponents. Here, it is assumed that the clusters that are formed are spherical ($d_f=d$) and that  the velocities of the constituent particles of a given cluster are uncorrelated implying that $\eta=1$. Substituting these values into Eqs.~(\ref{eqn:thea_general}) and (\ref{eqn:thetabeta_foralpha}), we reproduce the results
\be
\theta_n^{\mathrm{mf}}= \theta_e^{\mathrm{mf}}= \frac{2 d}{d+2},
\ee
where the superscript $\mathrm{mf}$ denotes mean field. Note that the main simplifying assumption is that $\eta=1$. In one dimension $\eta$ continues to be $1$ as the order of particles is maintained and a cluster made up of $m$ initial neighboring particles will have uncorrelated velocities. However, $\eta$ need not be $1$ in higher dimensions.

We now summarize the scaling theory predictions for the models studied in this paper.
For model A, since particles are point-like objects we have $r \sim m^0$ or $d_f=\infty$. Similarly, in model C since clusters are spherical $d_f = d$, which is spatial dimension itself. We thus obtain
\begin{align}
\theta_n &= 
\begin{cases}
\frac{2}{2+\eta}, &\text{model A,}\\
\frac{2d_f}{2d_f-2+\eta d_f}, &\text{model B,}\\
\frac{2}{1+\eta}, &\text{model C,}
\end{cases}
\label{eqn:exponents_relations}
\end{align}
with $\theta_e=\eta\theta_n$.

It is useful to have a relation between $\theta_n$ and $\theta_e$ that does not involve $\eta$. This will enable us to verify scaling theory without having to numerically measure the different exponents. Eliminating $\eta$, we obtain
\begin{alignat}{2}
2 \theta_n + \theta_e=&2, &&\quad \text{model A,}\nonumber\\
\frac{2 \theta_n}{2 \theta_n + \theta_e -2}=&d_f, &&\quad \text{model B,}
\label{eqn:hyperscaling}\\
\theta_n + \theta_e=&2, &&\quad \text{model C.}\nonumber
\end{alignat}

\section{\label{sec:results}Results}

In this section, we describe the results, obtained from extensive Monte Carlo simulations, for models $A$, $B$, and $C$. For all the three models, we will independently determine the exponents $\theta_n$, $\theta_e$, $\eta$ and $\zeta$. For model $B$ the fractal dimension $d_f$ is also measured. Their dependence on number density, the scaling relations between them, as well as deviation from the mean field results, are determined.

\subsection{\label{sec:results_mod1} Model A: Point particles}

\begin{figure}
  \includegraphics[width=\columnwidth]{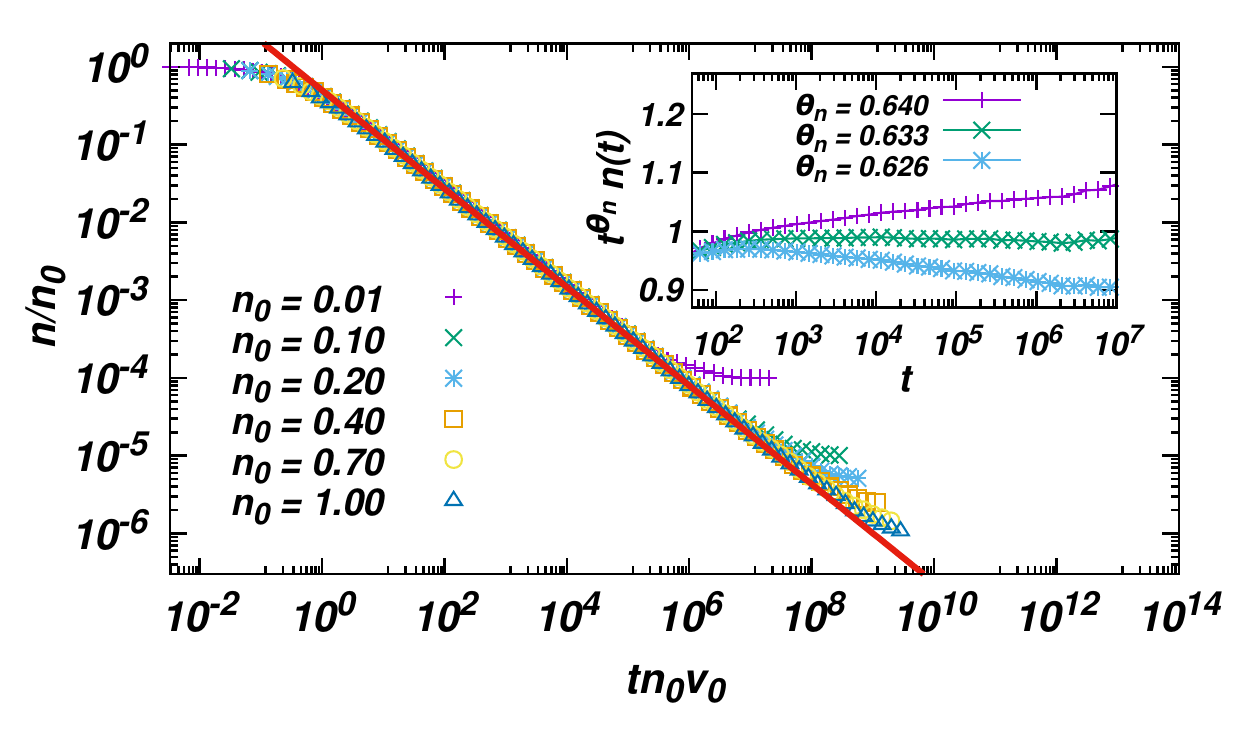}
  \caption{\label{fig:mod1tn} The data (model A) for mean number density of particles, $n(t)$, for different initial number densities $n_0$ collapse onto a single curve when $n(t)$ and $t$ are scaled as in Eq.~(\ref{eq:ntscaling}). The solid line is a power law $t^{-0.633}$. Inset: The compensated data  $n(t)t^{\theta_n}$ is shown for three different choices of $\theta_n$ differing by $0.007$ for $n_0=1.0$. The curve is flat for $\theta_n=0.633$.   The data are obtained for $L=1000$. All data have been averaged over 300 different initial conditions.}
\end{figure}

We first determine $\theta_n$ from the power law decay of the mean density of particles, $n$, with time $t$.
The data for different initial number density $n_0$ and initial mean speed $v_0$ collapse onto one curve when scaled, based on dimensional analysis, according to
\be
n(t,n_0)\simeq n_0 f(t n_0 v_0),
\label{eq:ntscaling}
\ee 
as shown in Fig.~\ref{fig:mod1tn}.  After an initial crossover time $t_c \sim n_0^{-1}$, $n(t)$ decreases as a power law. From the excellent collapse of the data for different $n_0$ onto one curve, we conclude that the power law  exponent is independent of the initial number density. From fitting a power law to the data,  we obtain $\theta_n = 0.633(7)$, which describes the data well over 5 decades. In the inset of Fig.~\ref{fig:mod1tn}, the compensated curve $t^{\theta_n} n(t)$  is shown for $n_0=1$. The mean slope of the curve changes from negative to positive as $\theta_n$ varies from $0.626$ to $0.640$, consistent with our  estimate of $\theta_n$ from direct measurement.

We now numerically determine $\theta_n$ using different analyses, both for the sake of consistency  as well as for benchmarking different methods that will be more useful in determining exponents for models B and C.

\begin{figure}
  \includegraphics[width=\columnwidth]{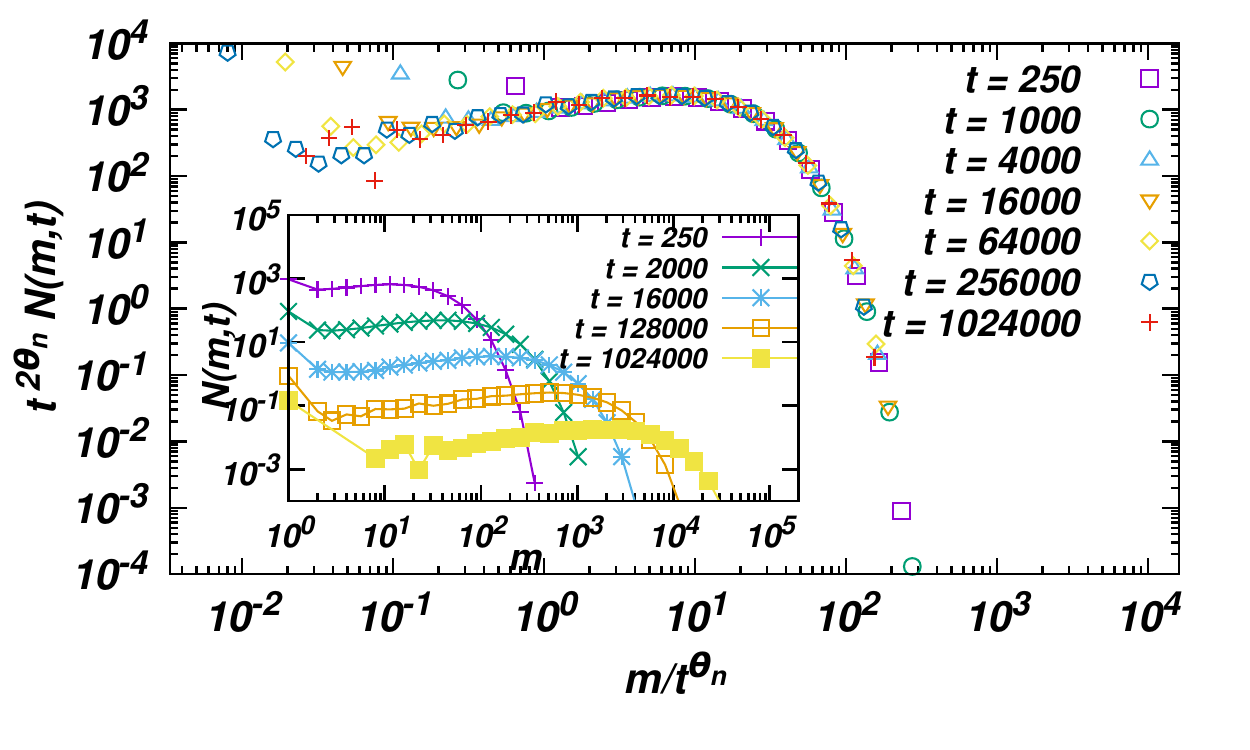}
  \caption{\label{fig:mod1nmt} The mass distribution ${N}(m,t)$ for different times collapse onto a single curve when scaled as in Eq.~(\ref{eqn:N(m,t)_final}), with $\theta_n=0.633$. The data are for  model A, with initial number density $n_0 = 1.0$, and system size $L=1000$ lattice. Inset: The unscaled data for $N(m,t)$ for different times $t$.}
\end{figure}

First we check that the measured value of $\theta_n$ is consistent with the mass distribution $N(m,t)$ and then we use the  finite size scaling for large times. The dependence of $N(m,t)$ on time and mass are shown in the inset of Fig.~\ref{fig:mod1nmt}. When scaled as in Eq.~(\ref{eqn:N(m,t)_final}) with $\theta_n=0.633$, the data for different times, that span three decades, collapse onto a single curve (see Fig.~\ref{fig:mod1nmt}). 

\begin{figure}
  \includegraphics[width=\columnwidth]{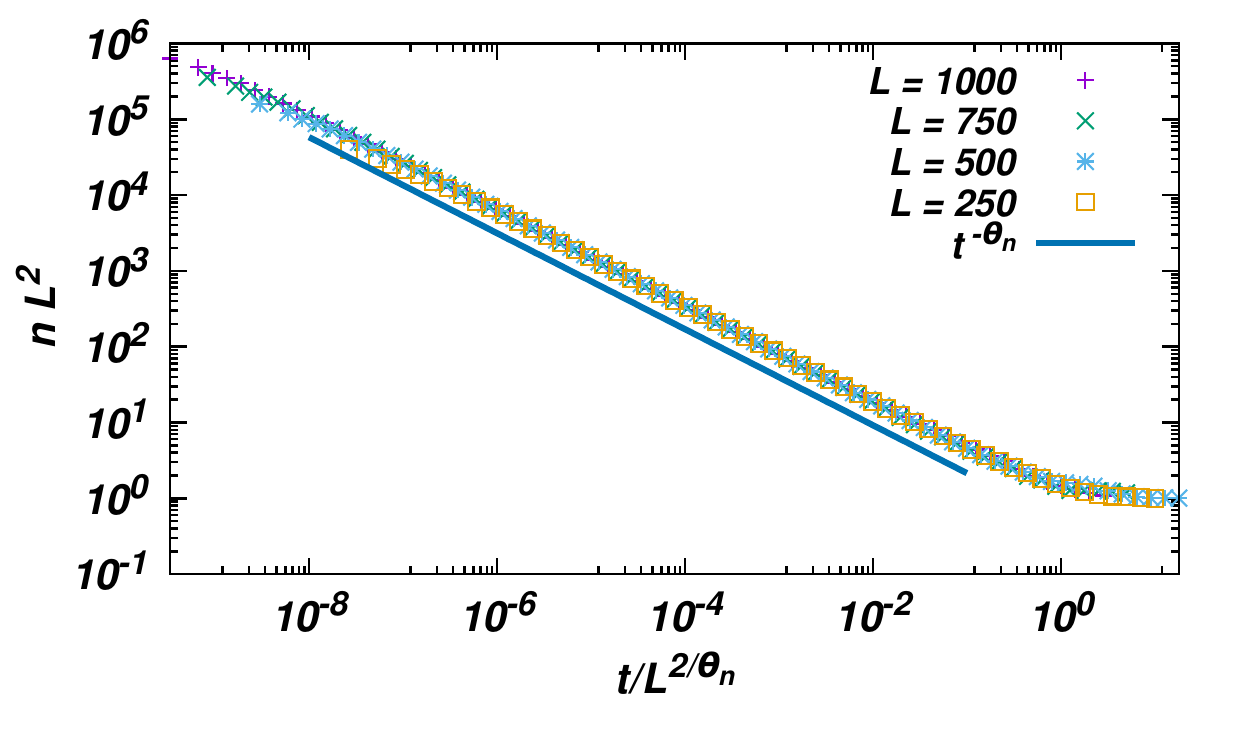}
  \caption{\label{fig:mod1_fse_n} 
The number density $n(t)$  for different system sizes $L$ collapse onto a single curve when scaled as in Eq.~(\ref{eqn:mod1_finitesize_n}), with $\theta_n=0.633$. The data are for  model A, and initial number density $n_0=1.0$.}
\end{figure}

Finally, we examine finite size effects. For very large times, when the number of clusters is order one, we expect that $n(t)\sim L^{-2}$, where $L$ is the system size. Assuming finite size scaling, we can write
\be
n(t) \simeq \frac{1}{L^2}f_n\left(\frac{t}{L^{2/\theta_n}}\right),
\label{eqn:mod1_finitesize_n}
\ee
where the scaling function $f_n(x) \sim x^{-\theta_n}$ for $x \ll 1$, and $f_n(x) \sim \text{constant}$ for $x \gg 1$.
The data for $n(t)$ for different $L$, when scaled  as in Eq.~(\ref{eqn:mod1_finitesize_n}) with $\theta_n = 0.633$, collapse onto a single curve, as shown in Fig.~\ref{fig:mod1_fse_n}. For model B and C, we will find the analysis of the data based on $N(m,t)$ and finite size scaling very useful for determining the exponents.

\begin{figure}
  \includegraphics[width=\columnwidth]{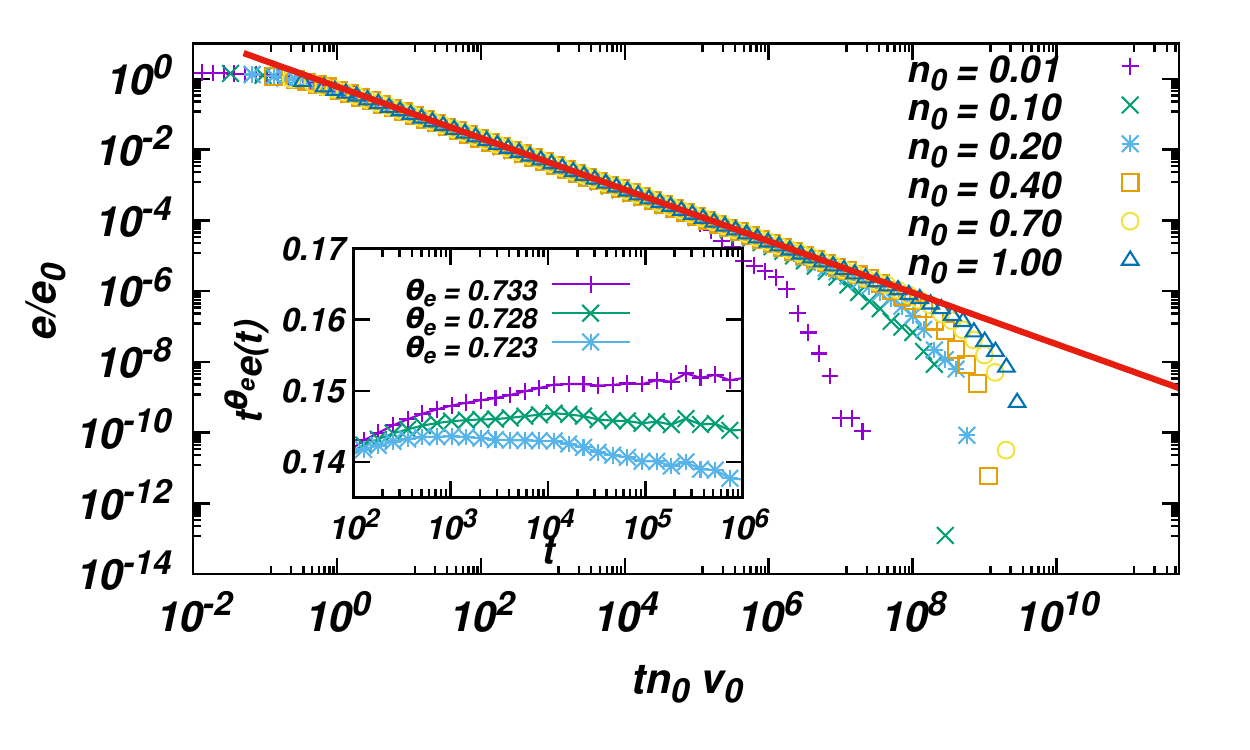}
  \caption{\label{fig:mod1tE} The data for mean energy density, $e(t)$, at time $t$ for different initial number densities $n_0$ in model A collapse onto a single curve when $e(t)$ and $t$ are scaled as shown in figure. The solid line is a power law $t^{-0.728}$. Inset: The compensated data  $n(t)t^{\theta_e}$ is shown for three different choices of $\theta_e$ differing by $0.005$ for $n_0=1.0$. The curve is flat for $\theta_e=0.728$.   The data are obtained for $L=1000$. All data have been averaged over 300 different initial conditions.
}
\end{figure}

We now determine $\theta_e$ from the power law decay of the mean energy density  $e$ with time $t$. 
The data for energy for different initial number density $n_0$, initial speed $v_0$ and initial mean energy $e_0$ collapse onto one curve when scaled, based on dimensional analysis, as
$e(t)\simeq e_0 f_e(t n_0 v_0)$,  as can be seen in Fig.~\ref{fig:mod1tE}.  After an initial crossover time $t_c \sim n_0^{-1}$, $e(t)$ decreases as a power law. From the excellent data collapse, we conclude that the power law  exponent is independent of the initial number density. From fitting a power law to the data,  we obtain $\theta_e = 0.728(5)$, which describes the data well over 5 decades. In the inset of Fig.~\ref{fig:mod1tE}, the compensated curve $t^{\theta_e} e(t)$  is shown for $n_0=1.0$. The mean slope of the curve changes from negative to positive as $\theta_n$ varies from $0.723$ to $0.733$, consistent with our direct measurement of $\theta_e$.

\begin{figure}
  \includegraphics[width=\columnwidth]{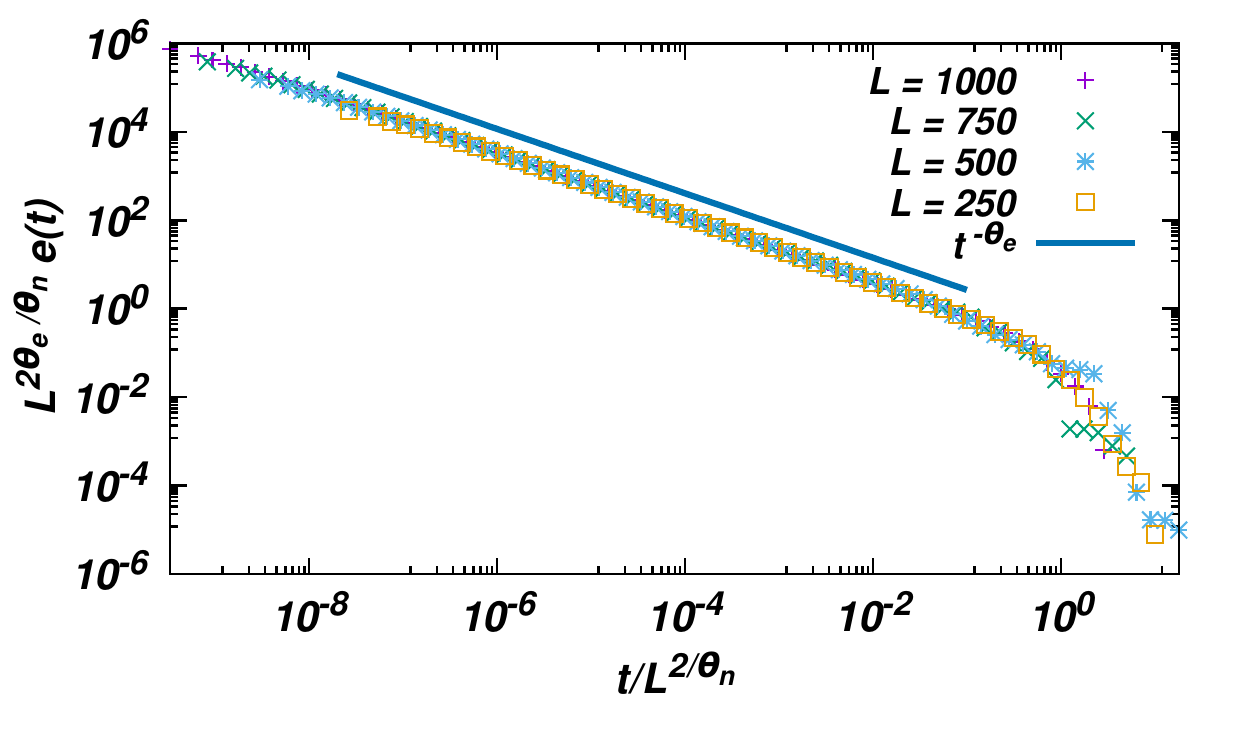}
  \caption{\label{fig:mod1_fse_E} 
The mean energy $e(t)$  for different system sizes $L$ collapse onto a single curve when scaled as in Eq.~(\ref{eqn:mod1_finitesize_E}), with $\theta_n=0.633$ and $\theta_e = 0.728$. The data are for  model A, and initial number density $n_0 = 1.0$.}
\end{figure}

The exponent $\theta_e$ can also be determined from finite size scaling.
As for number density,  $e(t)$ is expected to  obey finite size scaling of the form 
\be
e(t) \simeq \frac{1}{L^{2 \theta_e/\theta_n}}f_e\left(\frac{t}{L^{2/\theta_n}}\right),
\label{eqn:mod1_finitesize_E}
\ee
where the scaling function $f_e(x) \sim x^{-\theta_e}$ for $x \ll 1$, and $f_e(x) \sim \text{constant}$ for $x \gg 1$. 
The simulation data for different $L$ collapse onto a single curve (see Fig.~\ref{fig:mod1_fse_E}) when $e(t)$ and $t$ are scaled as in Eq.~(\ref{eqn:mod1_finitesize_E}) with $\theta_n = 0.633$ and $\theta_e = 0.728$. The power law extends over 4 decades.

\begin{figure}
  \includegraphics[width=\columnwidth]{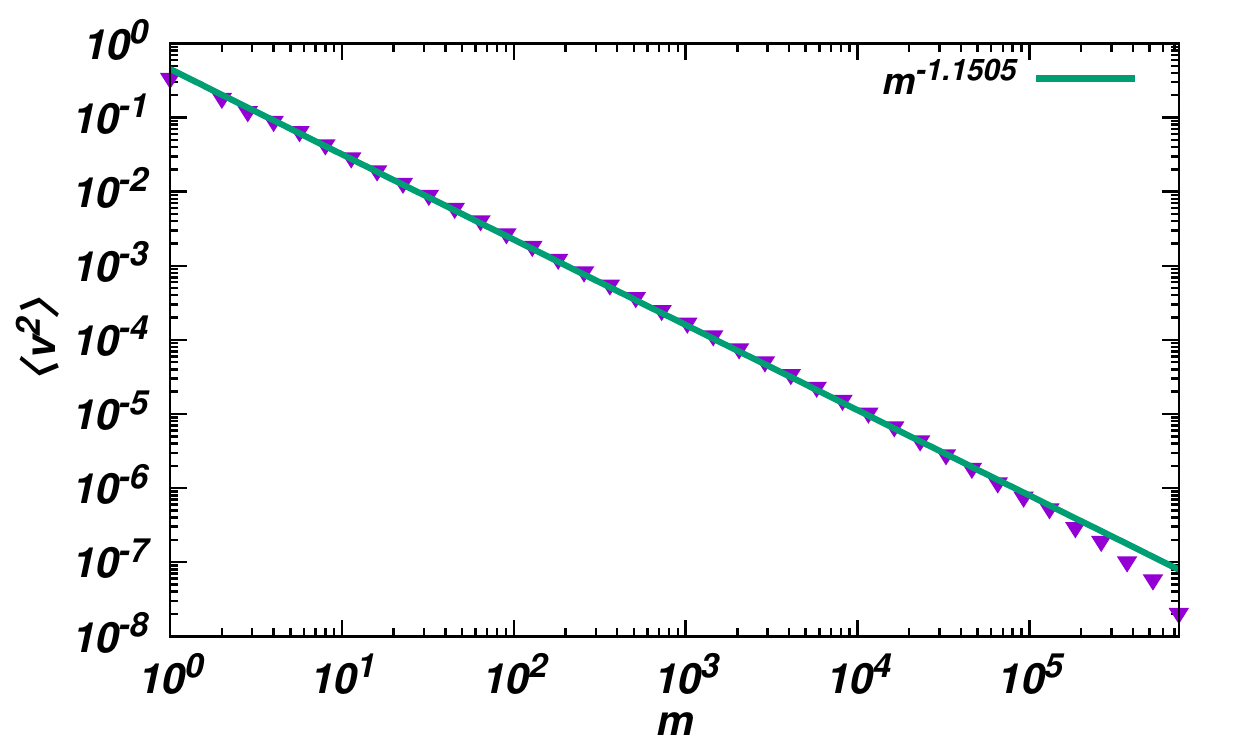}
  \caption{\label{fig:mod1vv} The variation of the mean square velocity $\langle v^{2} \rangle $ with mass $m$.  The solid line is power law $t^{-\eta}$ with $\eta=-1.1505$. The data are for model $A$, with $n_0 = 1.0$ and system size $L=1000$.}
\end{figure}

We now determine the exponent $\eta$ relating the scaling of velocity with mass as $v_m^2\sim m^{-\eta}$ [see Eq.~(\ref{eq:eta-defn})]. As seen from Fig.~\ref{fig:mod1vv}, $\langle v^2 \rangle$ for a fixed mass scales as a power law with $m$. We obtain  $\eta = 1.1505(3)$. 

Note that $\eta$ is not an independent exponent, but related to $\theta_n$ and $\theta_e$ through scaling theory, to be $\eta=\theta_e/\theta_n$ [see Eq.~(\ref{eqn:exponents_relations})]. From the measured values of $\theta_e=0.728$ and $\theta_n =0.633$, we obtain $\eta = 1.15$, consistent with the value from direct measurement $\eta=1.1505(3)$, thus providing support for the correctness of scaling theory.

\begin{figure}
  \includegraphics[width=\columnwidth]{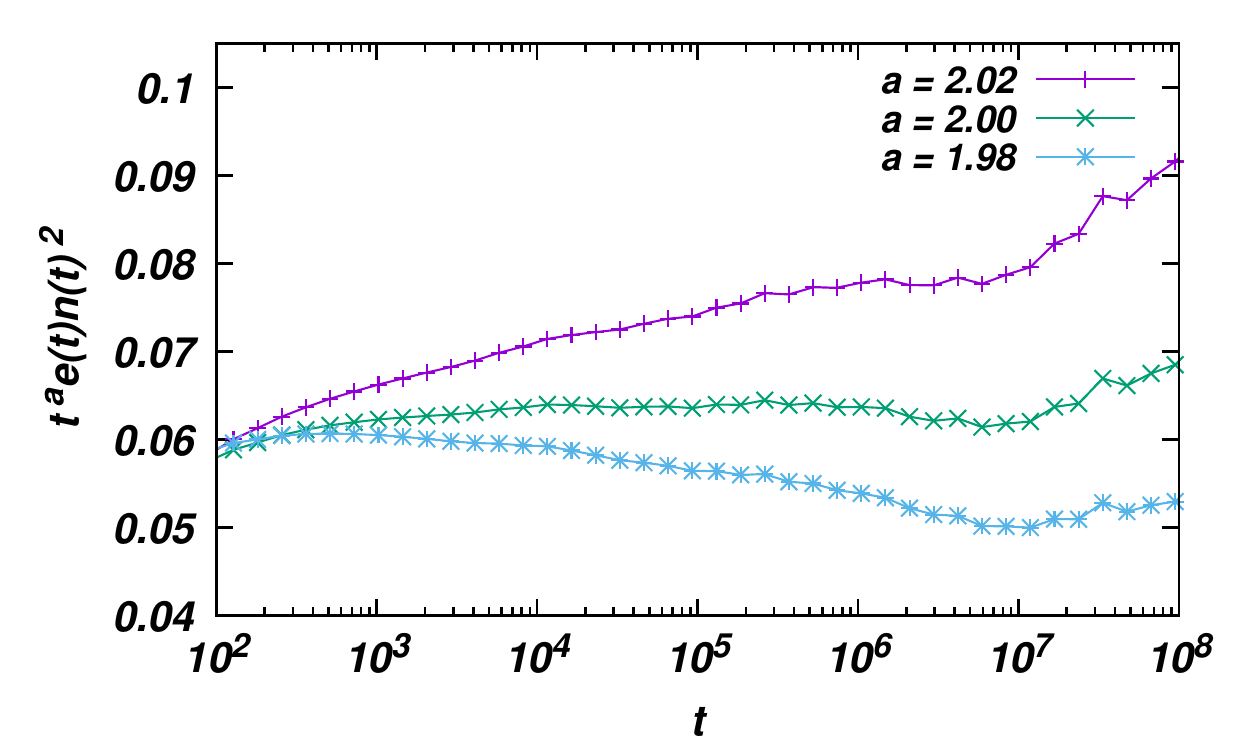}
  \caption{\label{fig:mod1_tenn}The variation of $t^a n^2(t)e(t)$ with time $t$ for three different values of $a$ close to $2$. The compensated curve is horizontal for $a=2.0$, validating the scaling relation in  Eq.~(\ref{eqn:hyperscaling}). The data are for model A, with initial number density $n_0 = 1.0$ and system size $L = 1000$.}
\end{figure}

We now provide a more direct evidence of scaling theory being correct. From Eqs.~(\ref{eqn:exponents_relations}) and (\ref{eqn:thetabeta_foralpha}), we obtain, by eliminating $\eta$, a relation between $\theta_e$ and $\theta_n$ as given in Eq.~(\ref{eqn:hyperscaling}).
If this relation is true, it implies that $t^2 n^2(t)e(t)$ should not depend on time $t$. In Fig.~\ref{fig:mod1_tenn}, we show the variation of $t^a n^2(t)e(t)$ with $a=1.98, 2.00, 2.02$. It is clear that only for $a=2.0$, the curve is horizontal. This gives us a way of validating the scaling relations without the need to measure any exponent directly.

\begin{figure}
  \includegraphics[width=\columnwidth]{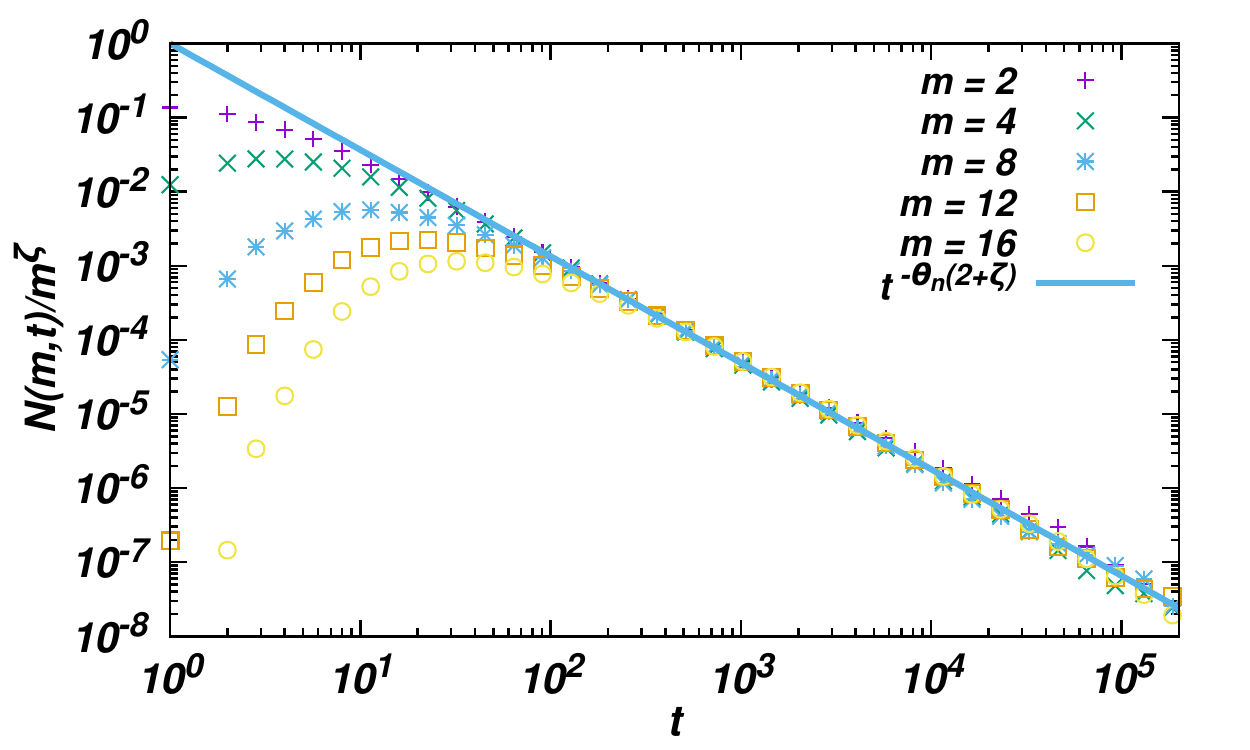}
  \caption{\label{fig:mod1_zeta}The data for $N(m,t)$ for different masses for large times collapse onto one curve when the number density is scaled as $N(m,t)/m^\zeta$ with $\zeta=0.270$. The solid line is a power law $t^{-\theta_n(2+\zeta)}$ with $\theta_n = 0.633$. The data are for model A, with initial number density $n_0=1.0$. }
\end{figure}

Finally, we determine the exponent $\zeta$ defined in Eq.~(\ref{eqn:small_mass_Nmt}) for small masses: $N(m,t) \sim m^\zeta t^{-\theta_n(2+\zeta)}$. Note that $\zeta$ is not related to $\theta_n$ or $\theta_e$ and is an independent exponent. To determine $\zeta$, we study the temporal behavior of $N(m,t)$ for fixed mass $m=2, 4, 8, 12, 16$. As shown in Fig.~\ref{fig:mod1_zeta}, the data for the different masses for large times collapse onto one curve when $N(m,t)$ is scaled as $N(m,t)/m^\zeta$, with $\zeta=0.270(5)$. We additionally check that the scaled data are consistent with the power law $t^{-\theta_n(2+\zeta)}$ for large times.

The numerically obtained values of the exponents for model A are summarized in Table~\ref{table_mod1}.
\begin{table}
\caption{\label{table_mod1}Summary of the numerically obtained values of the exponents for model A. The values are independent of initial density $n_0$.}
		\begin{ruledtabular}
		\begin{tabular}{c c}
		exponent & value\\
		\hline
		$\theta_n$ & 0.633(7)\\
		$\theta_e$ & 0.728(5)\\
		$\eta$ & 1.1505(3)\\
		$\zeta$ & 0.270(5)\\
	\end{tabular}
	\end{ruledtabular}
\end{table}

\subsection{Model B: Fractal Clusters} 

\begin{figure}
  \includegraphics[width=\columnwidth]{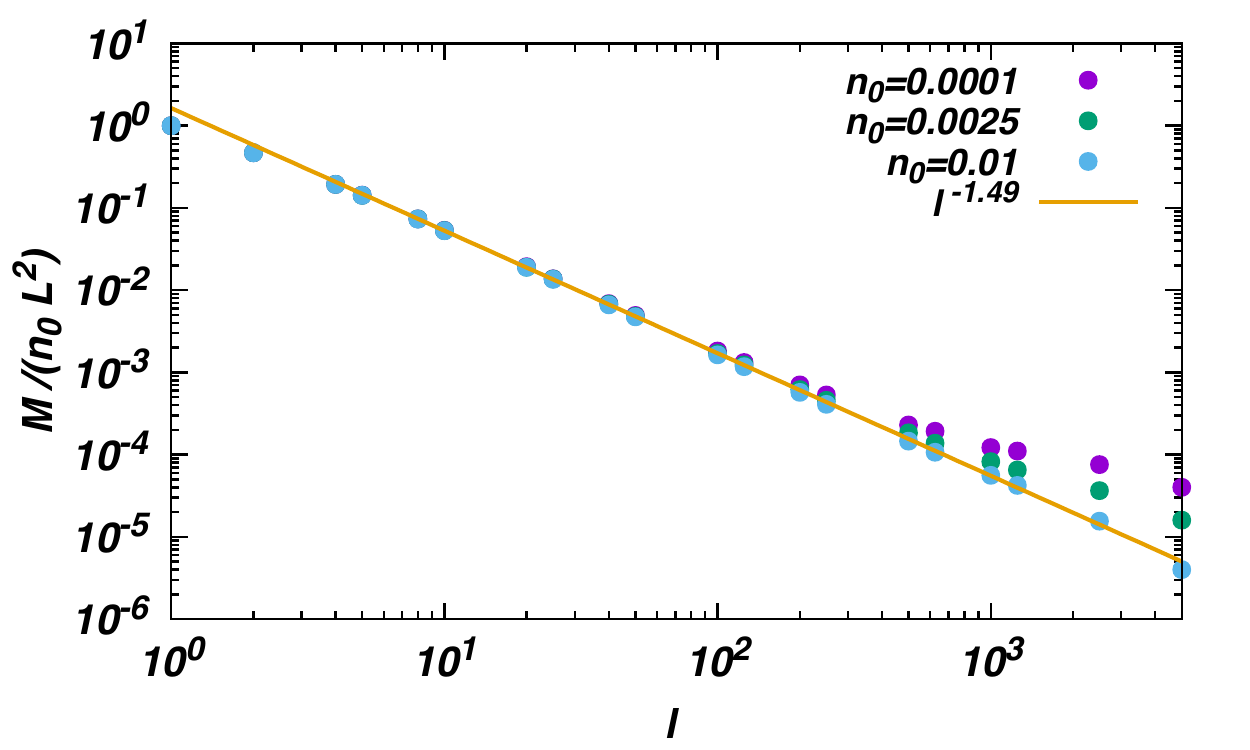}
  \caption{\label{fig:mod2_fracdim_0.0001} Determination of the fractal dimension of the largest cluster in model B using the box counting method. The number of non-empty boxes, $M$, varies with the size $\ell$ of the boxes used to tile the lattice as $M \sim \ell^{-d_f}$. We find $d_f\approx 1.49(3)$ (power law shown by solid line) irrespective of the initial density. The data are for  $L = 5000$.}
\end{figure}

In this subsection, we determine the exponents $\theta_n$, $\theta_e$, $\eta$ and $\zeta$ for model B. We first show that the clusters in model B are fractal with a fractal dimension, $d_f$, that lies between 1 and 2. To determine $d_f$, we consider  the final cluster in each of the simulations for a given initial number density $n_0$. $d_f$ of this cluster is measured using the  box counting method~\cite{fractalbook}. In this method, the lattice is tiled with square boxes  of  length $\ell$. Let $M$ be the  number of non-empty boxes. Then  $M\sim \ell^{-d_f}$. The results for three different $n_0$ are shown in Fig.~\ref{fig:mod2_fracdim_0.0001}. The data for different $n_0$ fall on top of each other for intermediate box sizes. The same is true for other $n_0$ and we conclude that $d_f$ is independent of $n_0$. We estimate $d_f$ to be 1.49(3).

\begin{figure}
  \includegraphics[width=\columnwidth]{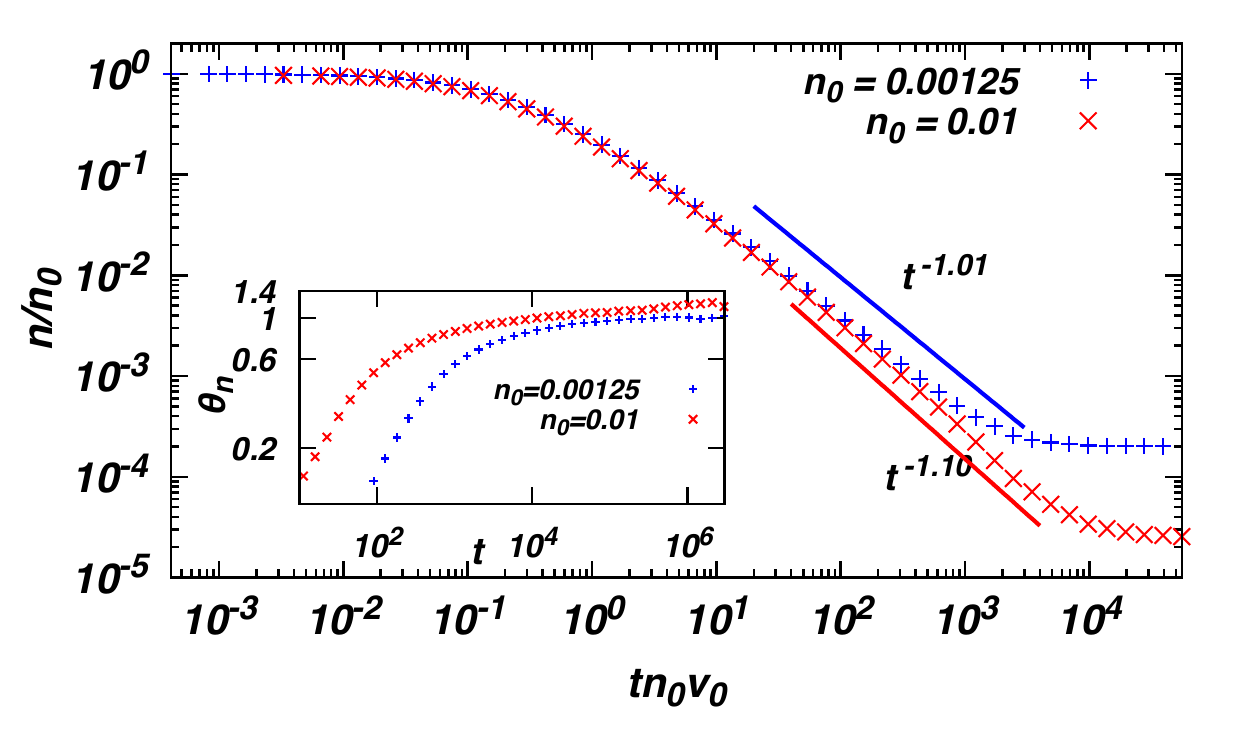}
  \caption{\label{fig:mod2_tn}The variation of the mean density of clusters $n(t)$ in model B with time $t$  is shown for two different initial densities. The exponents for the power laws, shown by solid lines,  have been obtained from finite size scaling. Inset: The time dependent exponent $\theta_n$ obtained from $ \theta_n=-d\ln n(t)/d\ln t$ is shown.  $\theta_n$ saturates for the larger initial densities only at late times. Data are for $L = 2000$ and averaged over $300$ different initial conditions. }
\end{figure}

Consider now the decay of the density of particles $n(t)$  with time $t$. We find that for model B, it is difficult to accurately determine $\theta_n$ directly from the data for $n(t)$ because of strong crossover effects. This can be seen from Fig.~\ref{fig:mod2_tn} where the variation of $n(t)$ with $t$ is shown  for two different initial densities $n_0=0.00125$ and $n_0=0.01$.  The data for the two densities overlap for short times but deviate for larger times. The solid lines, which are the estimates for $\theta_n$ from finite size scaling  (to be discussed below) match with the data only for late times. The convergence to the asymptotic answer can also be seen from measuring the instantaneous slope $ \theta_n=-d\ln n(t)/d\ln t$ for each time  (see inset of Fig.~\ref{fig:mod2_tn}). We find that the exponent $\theta_n$  saturates only  at late times for  the larger initial densities. We find that the same issue is present for the temporal decay of energy $e(t)$, making it also difficult to measure $\theta_e$ directly.

\begin{figure}
  \includegraphics[width=\columnwidth]{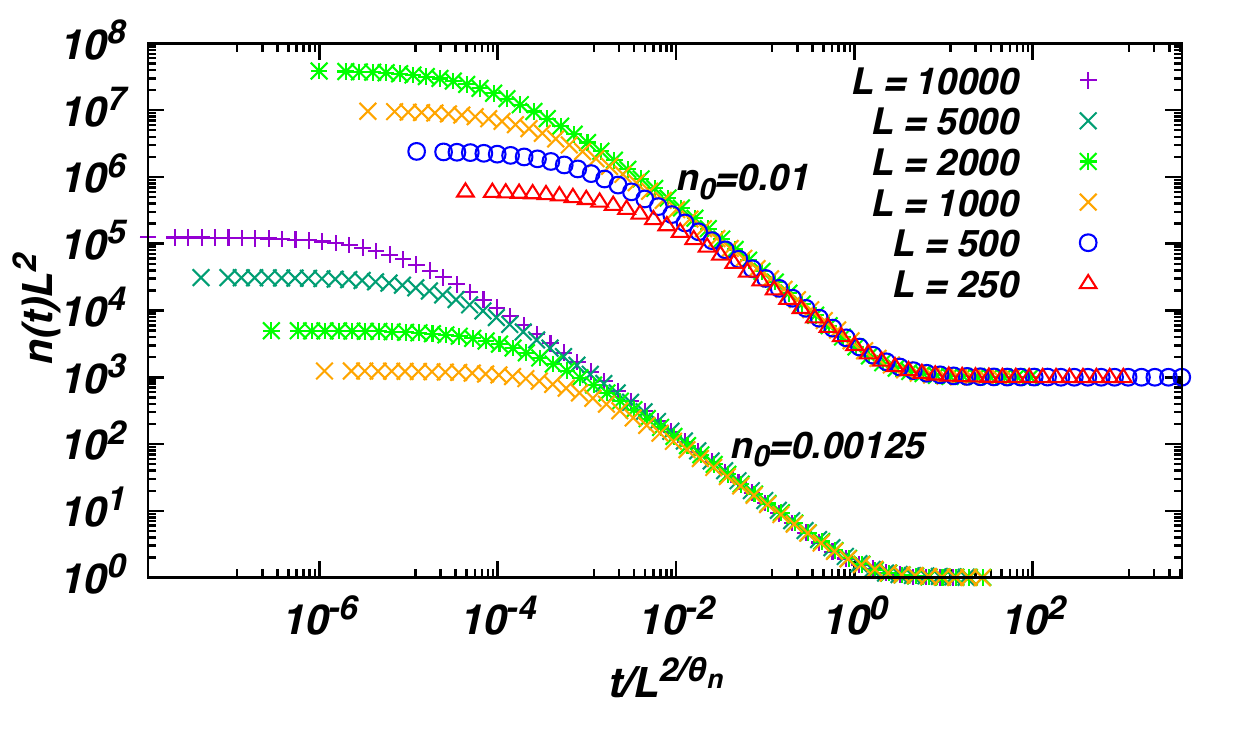}
  \caption{\label{fig:mod2_fse_n_125}Finite size scaling of $n(t)$ for model B: The number density $n(t)$  for different system sizes $L$ collapse onto a single curve when scaled as in Eq.~(\ref{eqn:mod1_finitesize_n}), with $\theta_n=1.01$ and $\theta_n=1.10$ for the initial densities $n_0 = 0.00125$ and $n_0 = 0.01$ respectively. The data for $n_0=0.01$ has been shifted for clarity.}
\end{figure}

We  determine $\theta_n$ from finite size scaling. For finite systems, $n(t)$ has the finite size scaling form given in Eq.~(\ref{eqn:mod1_finitesize_n}), namely $n(t) \simeq L^{-2} f_n(t/L^{2/\theta_n})$.  In Fig.~\ref{fig:mod2_fse_n_125}, we show the results for two representative initial densities $n_0=0.00125$ and $n_0=0.01$.  The data for different $L$, when scaled as in Eq.~(\ref{eqn:mod1_finitesize_n}), collapse onto a single curve with $\theta_n=1.01(1)$ for $n_0=0.00125$ and $\theta_n=1.10(1)$  for $n_0=0.01$. The results for other $n_0$ are listed in Table~\ref{table_mod2}, based on which  we conclude that $\theta_n$ depends on $n_0$ and converges to $\theta_n=1$ as $n_0 \to 0$. We also check that the same value of $\theta_n$ leads to the collapse of the data for $N(m,t)$ for different times when scaled as in Eq.~(\ref{eqn:N(m,t)_final}).

\begin{figure}
  \includegraphics[width=\columnwidth]{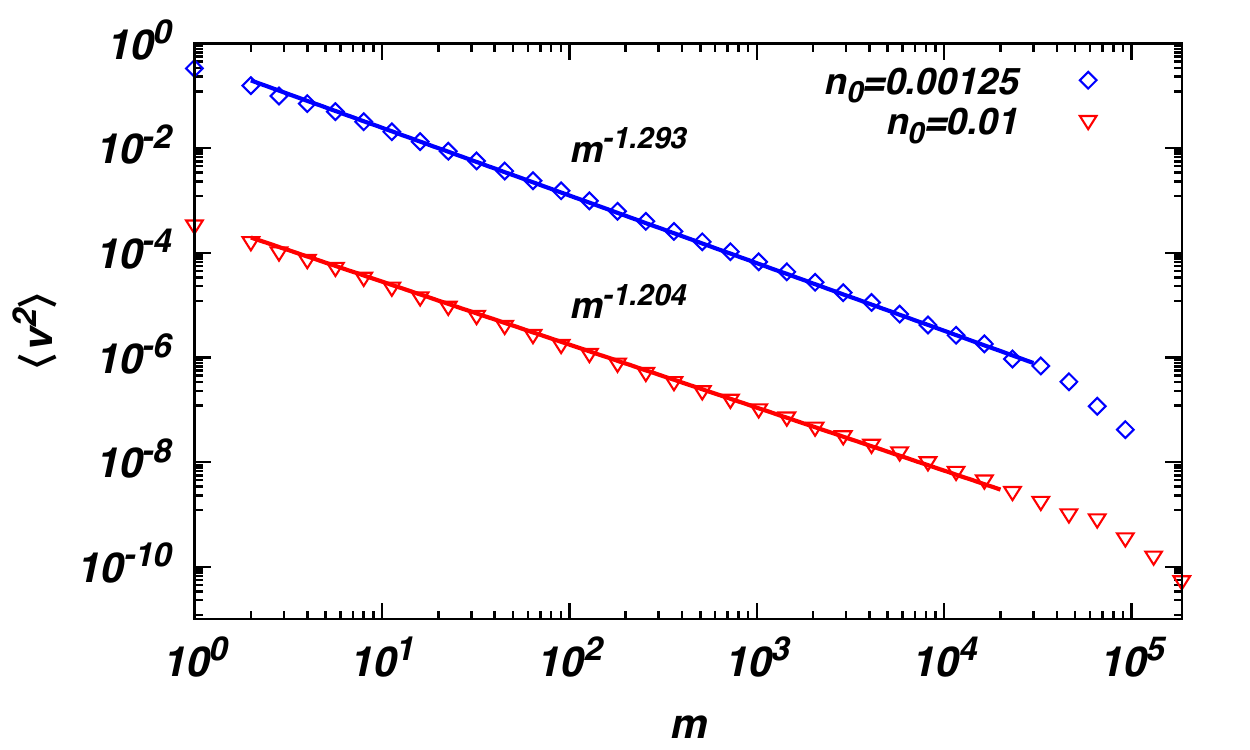}
  \caption{\label{fig:mod2_eta_0.125}The variation of the mean square velocity $\langle v^2 \rangle$ with mass $m$ for different initial densities. The solid lines are power-laws $m^{-\eta}$ with $\eta = 1.293(4)$ for $n_0 = 0.00125$ and $\eta=1.204(3)$ for $n_0 =0.01$. The data are for model B with system sizes $L = 10000$ for  $n_0 = 0.00125$ and $L=5000$ for $n_0 =0.01$. The data for $n_0=0.01$ has been shifted for clarity.}
\end{figure}

The limiting value of $\theta_n=1$ for $n_0 \to 0$ coincides with $\theta_n^{\mathrm{mf}}=1$. However, it is not clear whether the mean field result is obtained because correlations vanish. We check for correlations by measuring the exponent $\eta$. In Fig.~\ref{fig:mod2_eta_0.125}, we show the dependence of the mean squared velocity on the mass $m$ for two initial densities.  The power law dependence extends over three decades and we obtain exponents that depend on the initial density $n_0$ with  $\eta = 1.293(4)$ for $n_0=0.00125$ and $\eta=1.204(3)$ for  $n_0=0.01$. The results for other $n_0$ are listed in Table~\ref{table_mod2}, based on which  we conclude that $\eta$ also depends on $n_0$ and differs significantly from one for small $n_0$. However, as $n_0$ increases, we find that $\eta \to 1$. 

\begin{figure}
  \includegraphics[width=\columnwidth]{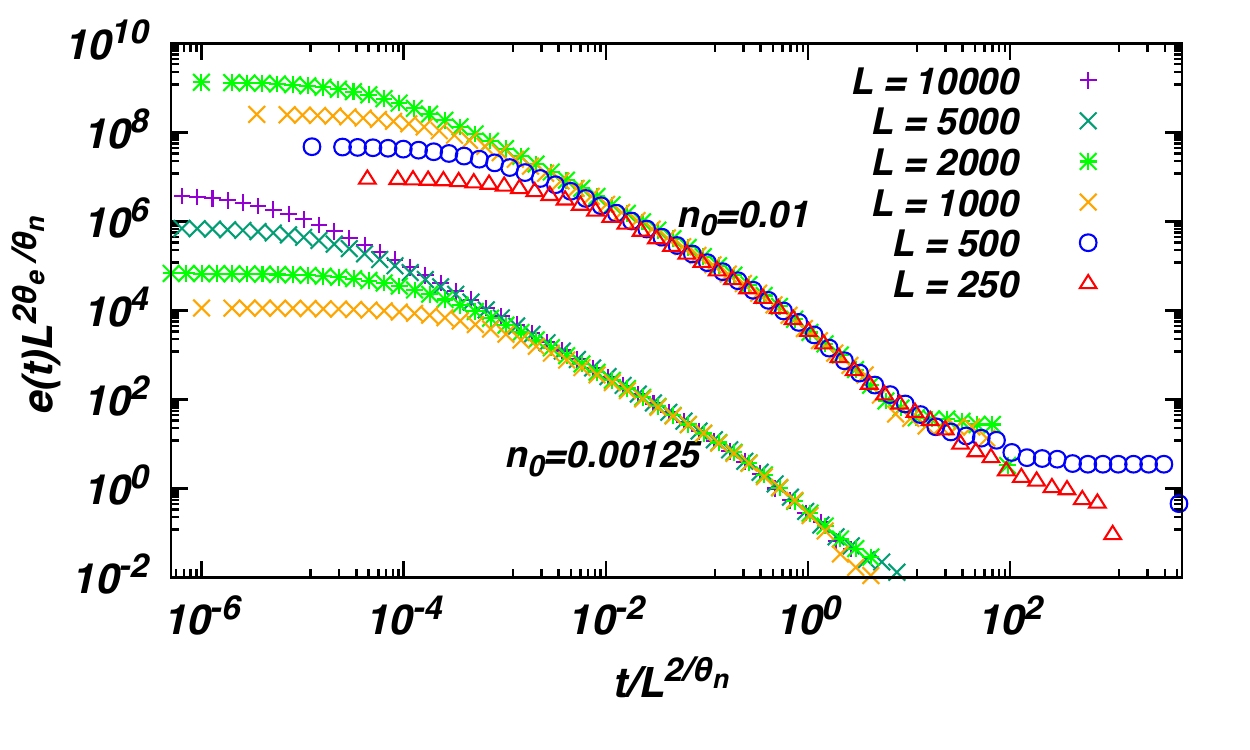}
  \caption{\label{fig:mod2_fse_e_125}Finite size scaling of $e(t)$ for model B: The mean energy $e(t)$  for different system sizes $L$ collapse onto a single curve when scaled as in Eq.~(\ref{eqn:mod1_finitesize_E}). Results for two different densities $n_0=0.00125$ and $n_0=0.01$(vertically shifted for visualization) is shown with $\theta_n$ obtained using finite size scaling [Eq.~(\ref{eqn:mod1_finitesize_n})] whereas $\theta_e$ is obtained using the hyperscaling relation [Eq.~(\ref{eqn:thetabeta_foralpha})].}
\end{figure}

Since it is difficult to measure $\theta_e$ directly from $e(t)$, we estimate $\theta_e$ using the scaling relation $\theta_e=\eta \theta_n$ [see Eq.~(\ref{eqn:thetabeta_foralpha})]. To check for consistency, we confirm that for this choice of $\theta_e$, the data for different system sizes  collapse onto one curve when $e(t)$ and $t$ are scaled using finite size scaling
 as described in Eq.~(\ref{eqn:mod1_finitesize_E}). The data collapse for two different $n_0$, shown in  Fig.~\ref{fig:mod2_fse_e_125}, is satisfactory. The results of $\theta_e$ for different $n_0$ are listed in Table~\ref{table_mod2}.

\begin{figure}
  \includegraphics[width=\columnwidth]{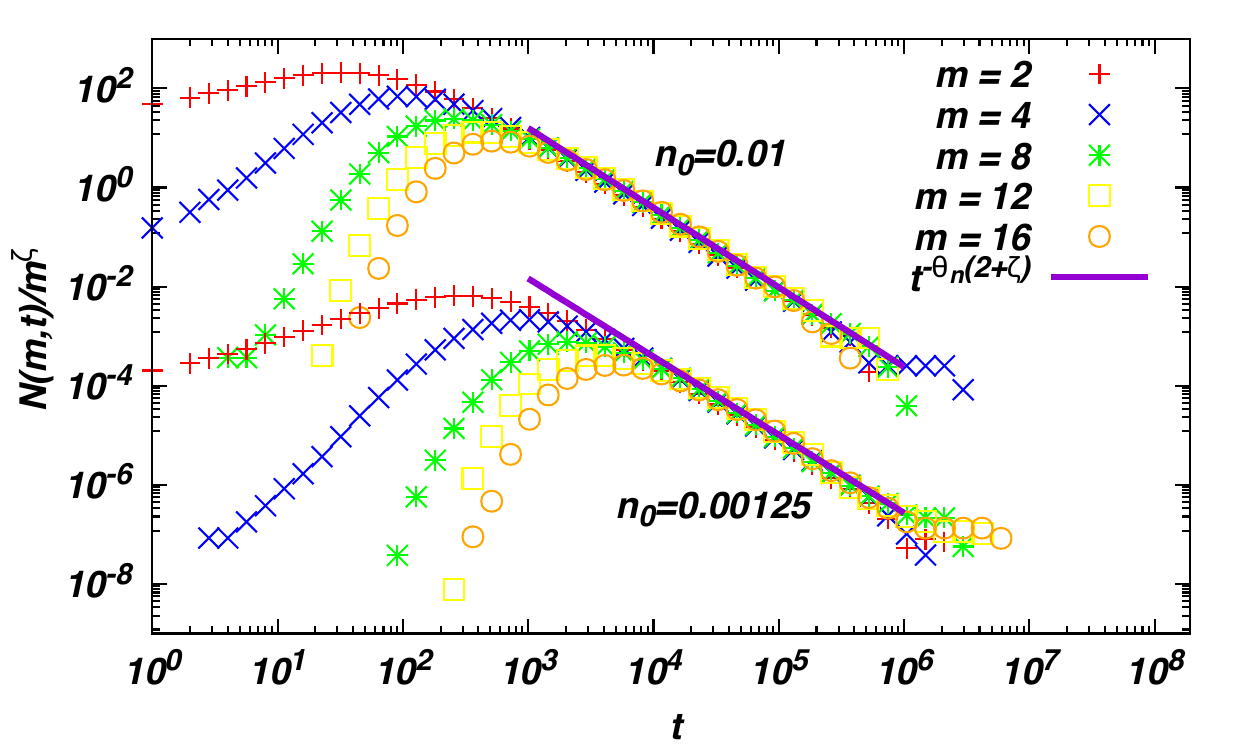}
  \caption{\label{fig:mod2_zeta}The data for $N(m,t)$ in model B for fixed masses collapse onto one curve when the number density is scaled as $N(m,t)/m^\zeta$ with $\zeta=-0.422(6)$ for $n_0=0.00125$ whereas $\zeta=-0.538(24)$ for $n_0=0.01$. The solid line is a power law $t^{-\theta_n(2+\zeta)}$ with $\theta_n$ taking values 1.01 and 1.10 for the initial densities 0.00125 and 0.01 (vertically shifted for visualization) respectively.}
\end{figure}

Finally, we determine the exponent $\zeta$ defined in Eq.~(\ref{eqn:small_mass_Nmt}) for small masses. Similar to model A, in order to determine $\zeta$, we study the temporal behavior of $N(m,t)$ for fixed mass $m=2, 4, 8, 12, 16$. Here, we illustrate the behavior of $\zeta$ for two different initial densities. As shown in Fig.~\ref{fig:mod2_zeta}, the data for the different masses collapse onto one curve for the respective initial densities when $N(m,t)$ is scaled as $N(m,t)/m^\zeta$, with $\zeta=-0.4229(6)$ for $n_0= 0.00125$ and $\zeta=-0.538(24)$ for $n_0= 0.01$. As an additional check, the scaled data are consistent with the power law with an exponent $t^{-\theta_n(2+\zeta)}$. 
Thus, the exponent $\zeta$ is dependent on the initial density $n_0$. Also, they are negative, as compared to model A where the exponent is positive.

\begin{figure}
\includegraphics[width  = \columnwidth]{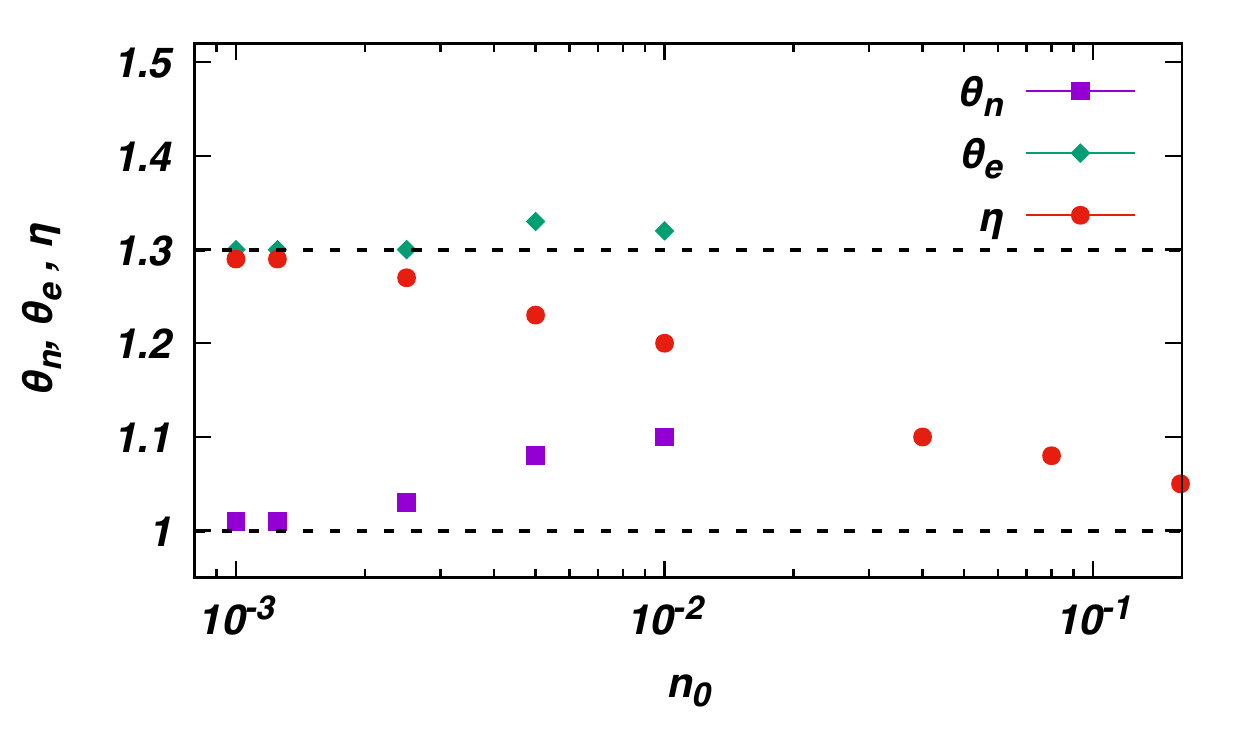}
\caption{\label{fig:mod2_exp_dendep} The variation of the exponents $\theta_n$, $\theta_e$ and $\eta$ are shown as function of the initial density $n_0$. The data are for model B. $\theta_n$ and $\theta_e$ approach an asymptotic limit $1.0$ and $1.3$ respectively for lowest densities. The exponent $\eta \approx 1.3$ in the low density limit and approaches the mean field result ($\eta=1$) for higher density.}
\end{figure}

The results for the exponents $\theta_n$, $\theta_e$, $\eta$, $d_f$ and $\zeta$ are summarized in Table~\ref{table_mod2} and their dependence on number density $n_0$ is shown in Fig.~\ref{fig:mod2_exp_dendep}. For higher densities, it is difficult to get the exponents $\theta _n$ and hence $\theta_e$ due to increasing finite-size effects. However, the exponent $\eta$ can be calculated for the densities larger than $0.01$. From Table~\ref{table_mod2}, we observe that, when $n_0 \to 0$, the exponents tend to the limiting values $\theta_n \to 1$, $\eta \to 1.3$ and $\theta_e \to 1.3$. When the density increases, we find that $\eta \to 1$, thus approaching its mean field value $\eta^{\mathrm{mf}} =1$. We conclude that velocity correlations vanish as density increases. We note that in model B, there are no avalanche of coalescence events caused due to two clusters colliding.  We also verify that the exponents satisfy the hyperscaling relation given by Eq.~(\ref{eqn:hyperscaling}). In Table~\ref{table_mod2}, the fractal dimension determined numerically is compared with that obtained by Eq.~(\ref{eqn:hyperscaling}) [see columns 5 and 6]. For all densities, the values are equal within error bars, thus consistent with the scaling theory.
\begin{table}
\caption{\label{table_mod2}Summary of the numerically obtained values of the exponents for model B.}
		\begin{ruledtabular}
		\begin{tabular}{ c c c c c c c }
		$n_0$  & $\theta_n$   &  $\eta$  & $\theta_e$  & $d_f$ & $d_f$ & $\zeta$ \\ 
		& & &{\small ($=\eta \theta_n$)}  & &{\small [Eq.~(\ref{eqn:hyperscaling})]} & \\
		\hline
		0.00100		&	1.01(5)				&	1.291(4) &  1.30(7)		&		1.49(3) &1.54(17)& -0.41(5) \\
		0.00125		&	1.01(8)				&	1.293(4) & 1.30(10)		&		1.49(3) & 1.54(23)& -0.42(1) \\
		0.00250		&	1.03(4)				&	1.261(7)	 & 1.30(6)	&		1.49(3) & 1.52(14)& -0.46(2)  \\
		0.00500		&	1.08(4)				&	1.231(2)  & 1.33(5)		&		1.49(3) & 1.46(12)& -0.49(2)  \\
		0.01			&	1.10(2)				&	1.204(3)  & 1.32(3)		&		1.49(3) & 1.45(7)& -0.54(2)  \\
		0.04			&	---				&	1.10	 & ---	&---&		--- & ---\\
		0.08			&	---				&	1.08	 & ---	&---&		--- & ---\\
		0.16			&	---				&	1.05	 & ---	&---&		---& ---\\
	\end{tabular}
	\end{ruledtabular}
\end{table}

\subsection{Model C: Spherical clusters}

\begin{figure}
	\includegraphics[width=\columnwidth]{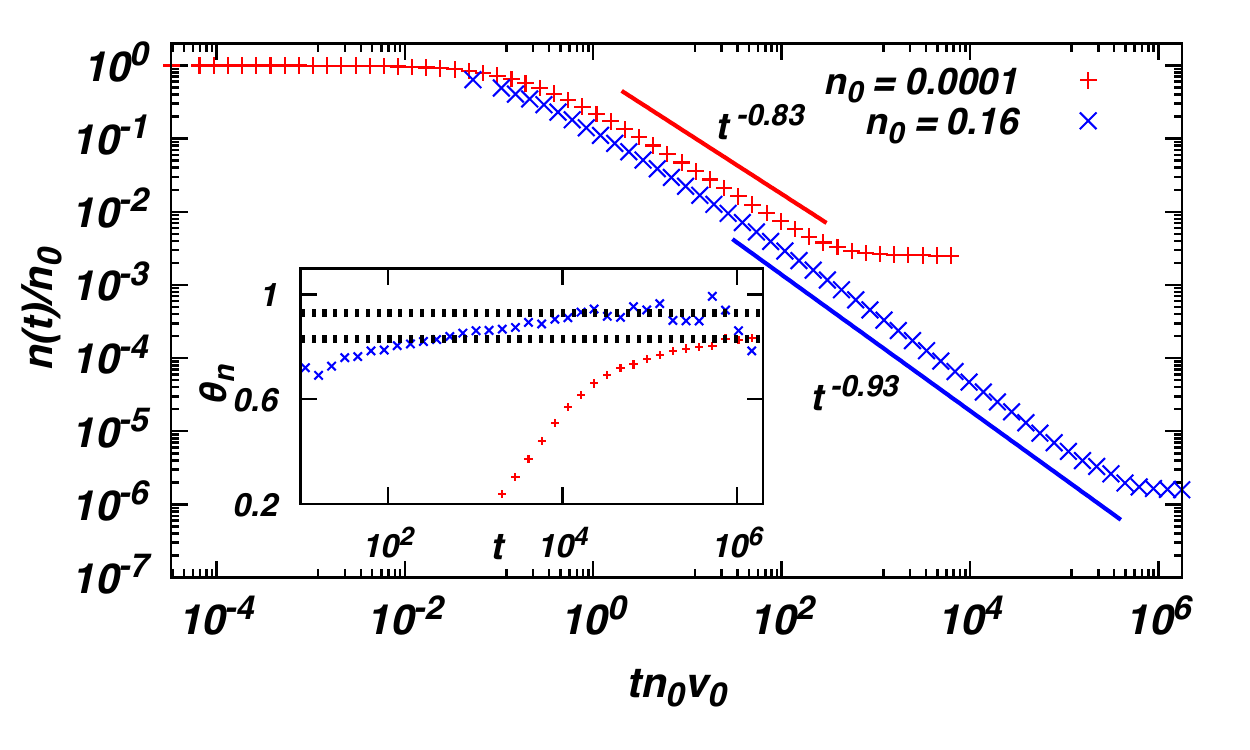}
	\caption{\label{fig:mod3_dendep_n}The variation of the mean density of clusters $n(t)$ in model C with time $t$  is shown for two different initial densities. The exponents for the power laws, shown by solid lines,  have been obtained from finite size scaling. Inset: The time dependent exponent $\theta_n$ obtained from $ \theta_n=-d\ln n(t)/d\ln t$ is shown.  $\theta_n$ saturates for the larger initial densities only at late times. The dashed lines are the reference for  the exponents 0.83 and 0.93. Data are for $L = 2000$ and averaged over $300$ different initial conditions.}
\end{figure}

We now determine the exponents $\theta_n$, $\theta_e$, $\eta$ and $\zeta$ for model C. We first show that the exponent $\theta_n$ depends on initial densities $n_0$. Figure~\ref{fig:mod3_dendep_n} shows the variation of $n(t)$ with time $t$ for two different initial densities, one small and one large. The time dependent $\theta_n=-d \ln n(t)/ d \ln t$, shown in the inset, saturates at different values for the different initial densities. Like for model B, it is difficult to measure $\theta_n$ directly as $n(t)$ shows strong crossover effects. For this reason, we determine $\theta_n$ from finite size scaling (see below) following which we obtain $\theta_n=0.83$ for $n_0=0.0001$ and $\theta_n=0.93$ for $n_0=0.16$. The exponents obtained from finite size scaling are shown in Fig.~\ref{fig:mod3_dendep_n} for comparison and they  describe the data for large times well. 

\begin{figure}
\includegraphics[width=\columnwidth]{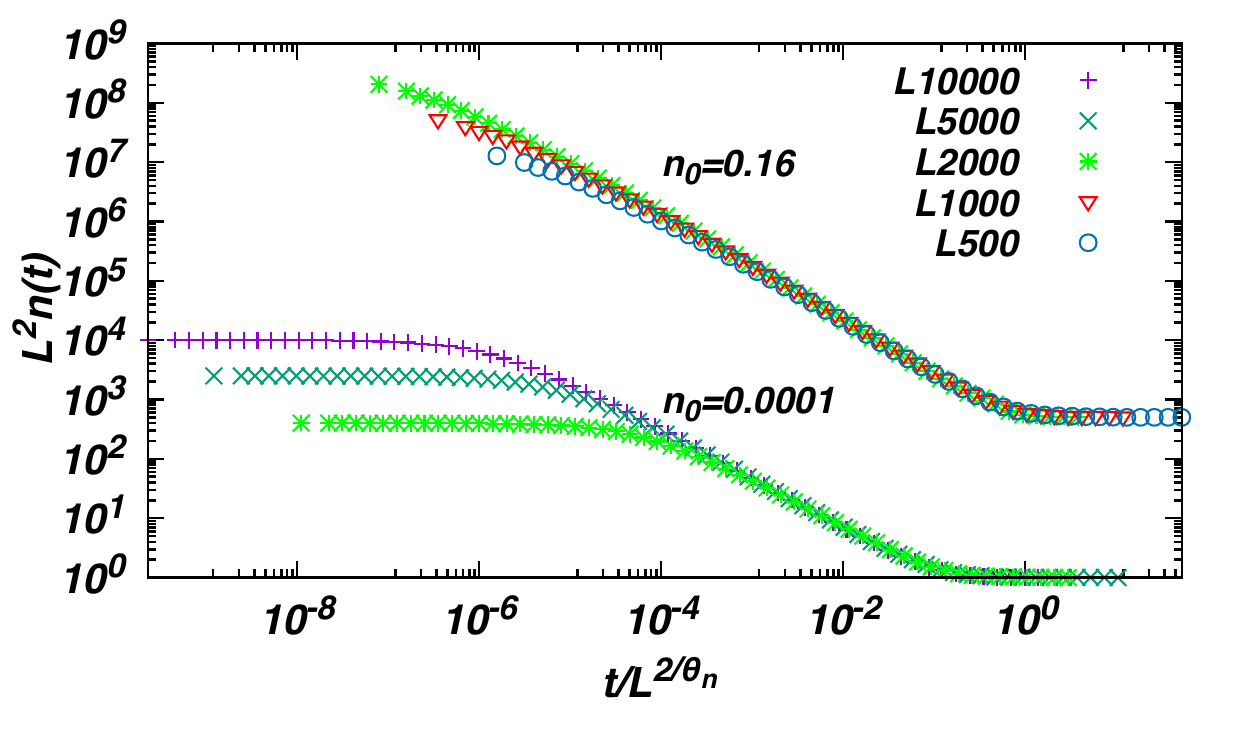}
\caption{\label{fig:mod3_fse_n(t)}Finite size scaling of $n(t)$ for model C: The number density $n(t)$  for different system sizes $L$ collapse onto a single curve when scaled as in Eq.~(\ref{eqn:mod1_finitesize_n}), with $\theta_n=0.83(4)$ and $\theta_n=0.93(5)$ for the initial densities $n_0 = 0.0001$ and $n_0 = 0.16$ respectively. The data for $n_0=0.16$ has been shifted for clarity.}
\end{figure}

We determine the exponent $\theta_n$ using the finite size scaling $n(t) \simeq L^{-2} f_n(t/L^{2/\theta_n})$ [see Eq.~(\ref{eqn:mod1_finitesize_n})].  Two representative cases are shown in Fig.~\ref{fig:mod3_fse_n(t)}. The data of $n(t)$ for different $L$, when scaled as in Eq.~(\ref{eqn:mod1_finitesize_n})  collapse onto a single curve for $\theta_n=0.83$  for $n_0=0.001$ and $\theta_n=0.93$  for $n_0=0.16$. The results for other $n_0$ are  listed in Table~\ref{table_mod3}, based on which we conclude that $\theta_n$  depends on $n_0$ and increases to the mean field result $\theta^{\mathrm{mf}}_n=1$ with increasing $n_0$. We also check that the same value of $\theta_n$ leads to the collapse of the data for $N(m,t)$ for different times when scaled as in Eq.~(\ref{eqn:N(m,t)_final}).

\begin{figure}
	\includegraphics[width = \columnwidth]{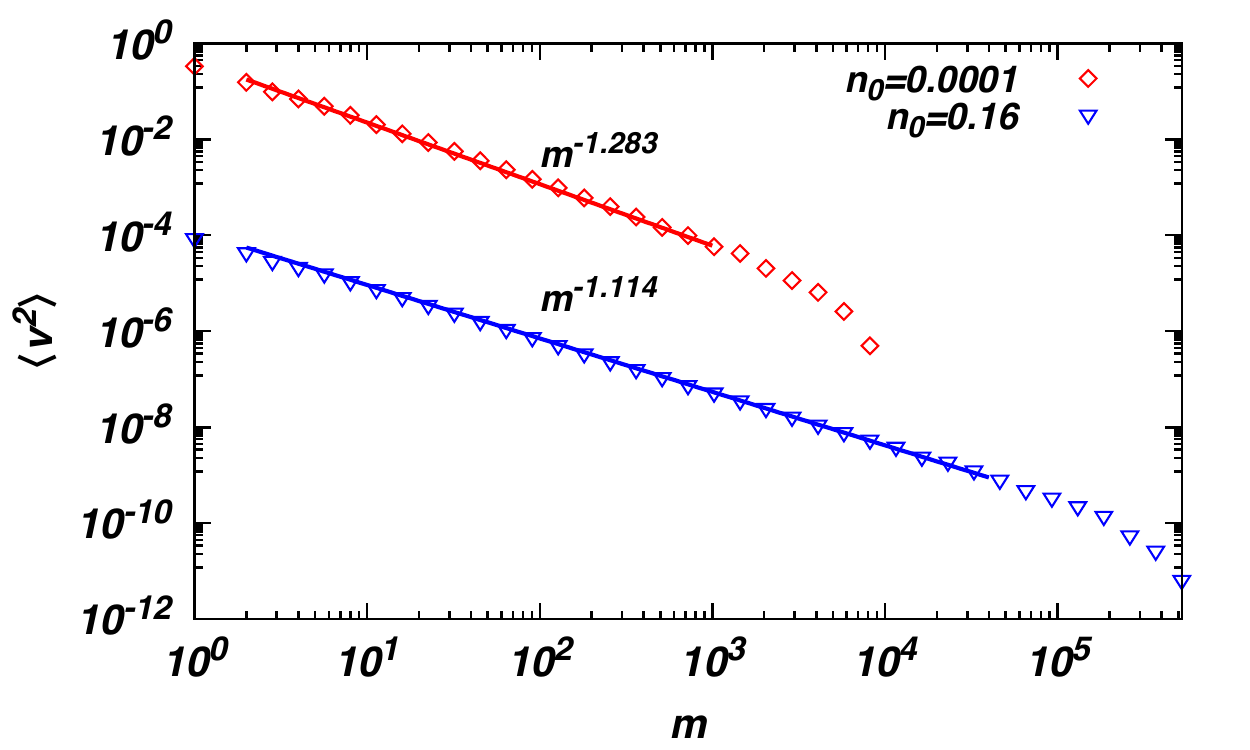}
	\caption{\label{fig:mod3_vv_0.01}The variation of the mean square velocity $\langle v^2 \rangle$ plotted as function of mass $m$ for different initial densities. The solid lines are power-laws $m^{-\eta}$ with $\eta = 1.283(13)$ and $\eta=1.114(2)$ for $n_0 = 0.0001$ and $n_0=0.16$ respectively. The data are for model C with system sizes $L = 10000$ and $L=2000$ for the densities $n_0 = 0.0001$ and $n_0=0.16$ respectively. The data for $n_0=0.16$ has been shifted for clarity.}
\end{figure}

It is possible that the mean field result is obtained at higher $n_0$ because the correlations vanish. Two representative cases are shown in Fig.~\ref{fig:mod3_vv_0.01}. We find that $\eta$ depends on  the initial density $n_0$ with $\eta = 1.283(13)$ for $n_0 = 0.0001$ and $\eta=1.114(2)$ for $n_0 = 0.16$.  The results for other $n_0$ are listed in Table~\ref{table_mod3} and it shows that $\eta$ decreases to its mean field prediction $\eta^{\mathrm{mf}}=1$ as density increases.
\begin{figure}
\includegraphics[width  = \columnwidth]{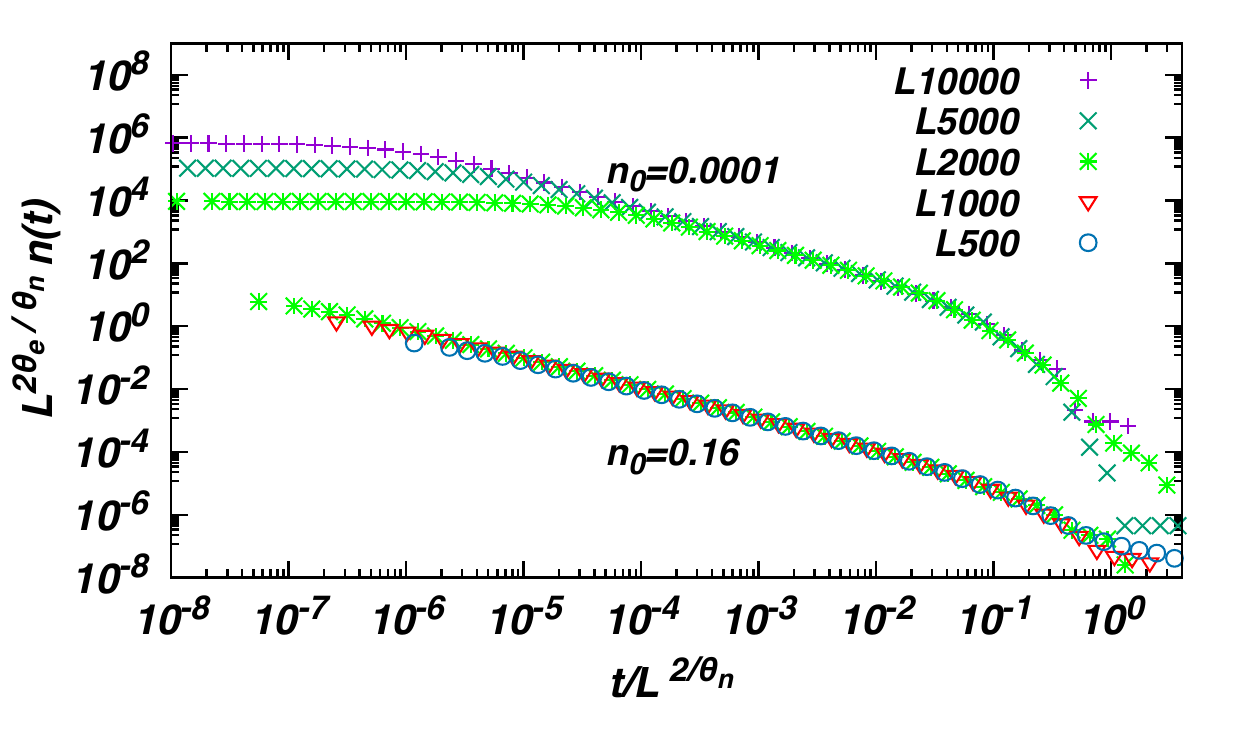}
\caption{\label{fig:mod3_fse_e(t)} Finite size scaling of $e(t)$ for model C: The mean energy density $e(t)$  for different system sizes $L$ collapse onto a single curve when scaled as in Eq.~(\ref{eqn:mod1_finitesize_E}). Results for two different densities $n_0=0.0001$ and $n_0=0.16$ is shown with $\theta_n$ obtained using finite size scaling [Eq.~(\ref{eqn:mod1_finitesize_n})] whereas $\theta_e$ obtained using the hyperscaling relation [Eq.~(\ref{eqn:thetabeta_foralpha})]. The data for $n_0=0.0001$ has been shifted for clarity.}
\end{figure}

We find that it is difficult to measure $\theta_e$ directly from the power-law decay of $e(t)$. Hence, we measure $\theta_e$ using the scaling relation, $\theta_e=\eta \theta_n$ [see Eq.~(\ref{eqn:thetabeta_foralpha})]. To check for the consistency of the result for $\theta_e$ obtained using the scaling relation~[Eq.~(\ref{eqn:thetabeta_foralpha})], we confirm that for this choice of $\theta_e$, the data for different system sizes can be collapsed onto one curve using finite size scaling $e(t) \simeq L^{-2 \theta_e/\theta_n} f_e(t/L^{2/\theta_n})$  [see Eq.~(\ref{eqn:mod1_finitesize_E})]. The data collapse is satisfactory as shown in Fig.~\ref{fig:mod3_fse_e(t)} for the two different $n_0$. The results of $\theta_e$ for different $n_0$ are listed in Table~\ref{table_mod3} which shows that $\theta_e$ is close to the mean field limit, $\theta^{\mathrm{mf}}_e=1$ for all $n_0$.

\begin{figure}
  \includegraphics[width=\columnwidth]{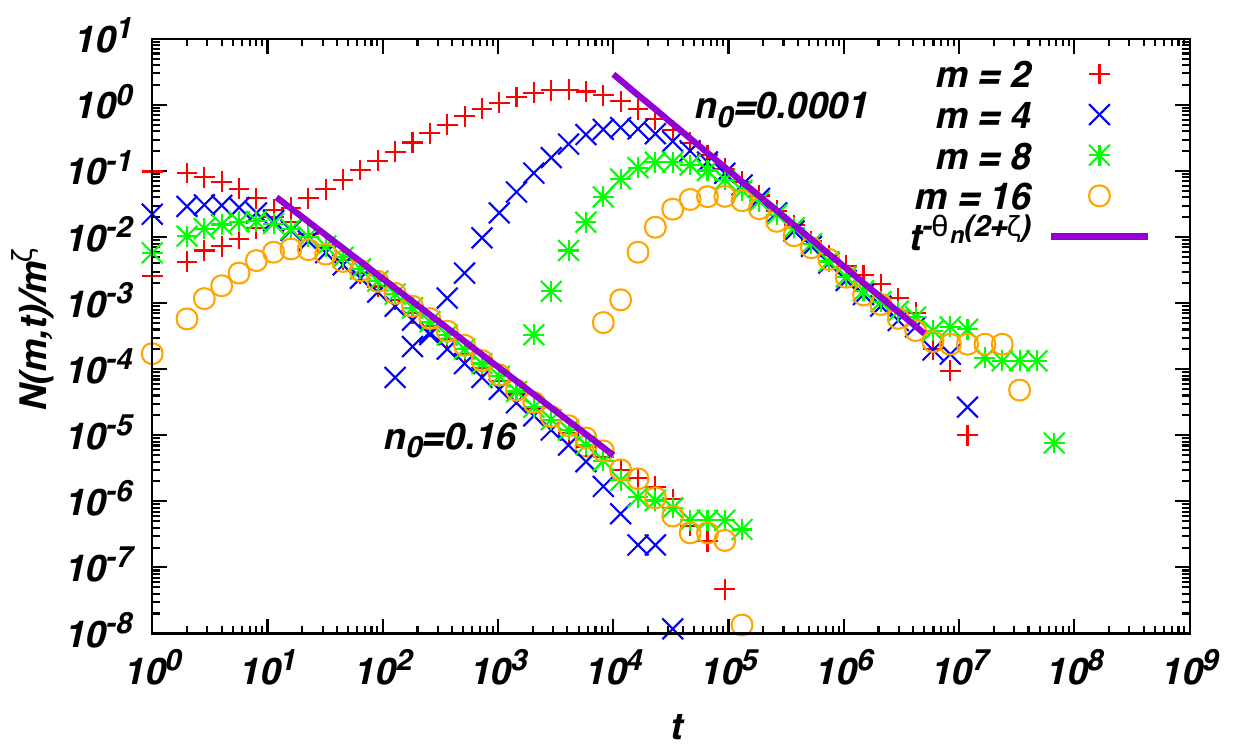}
  \caption{\label{fig:mod3_zeta}The data for $N(m,t)$ in model C for fixed masses collapse onto one curve when the number density is scaled as $N(m,t)/m^\zeta$ with $\zeta=-0.248(26)$ for $n_0=0.0001$ whereas $\zeta=-0.563(10)$ for $n_0=0.16$. The solid line is a power law $t^{-\theta_n(2+\zeta)}$ with $\theta_n$ as 0.83 and 0.93 for the initial densities 0.0001 and 0.16 respectively. The data for $n_0=0.0001$ has been shifted for clarity.}
\end{figure}

Finally, we determine the exponent $\zeta$ [defined in Eq.~(\ref{eqn:small_mass_Nmt})] for small masses. In order to determine $\zeta$, we study the temporal behavior of $N(m,t)$ for fixed mass $m=2, 4, 8, 16$. Here, we illustrate the behavior of $\zeta$ for two different initial densities. As shown in Fig.~\ref{fig:mod3_zeta}, the data for the different masses collapse onto one curve for the respective initial densities when $N(m,t)$ is scaled as $N(m,t)/m^\zeta$, with $\zeta=-0.248(26)$ for $n_0= 0.0001$ and $\zeta=-0.563(10)$ for $n_0= 0.16$. As an additional check, the scaled data are consistent with the power law with an exponent $t^{-\theta_n(2+\zeta)}$.  The results of $\zeta$ for other densities are listed in Table~\ref{table_mod3}. We conclude that $\zeta$ is strongly dependent on $n_0$.

\begin{figure}
\includegraphics[width  = \columnwidth]{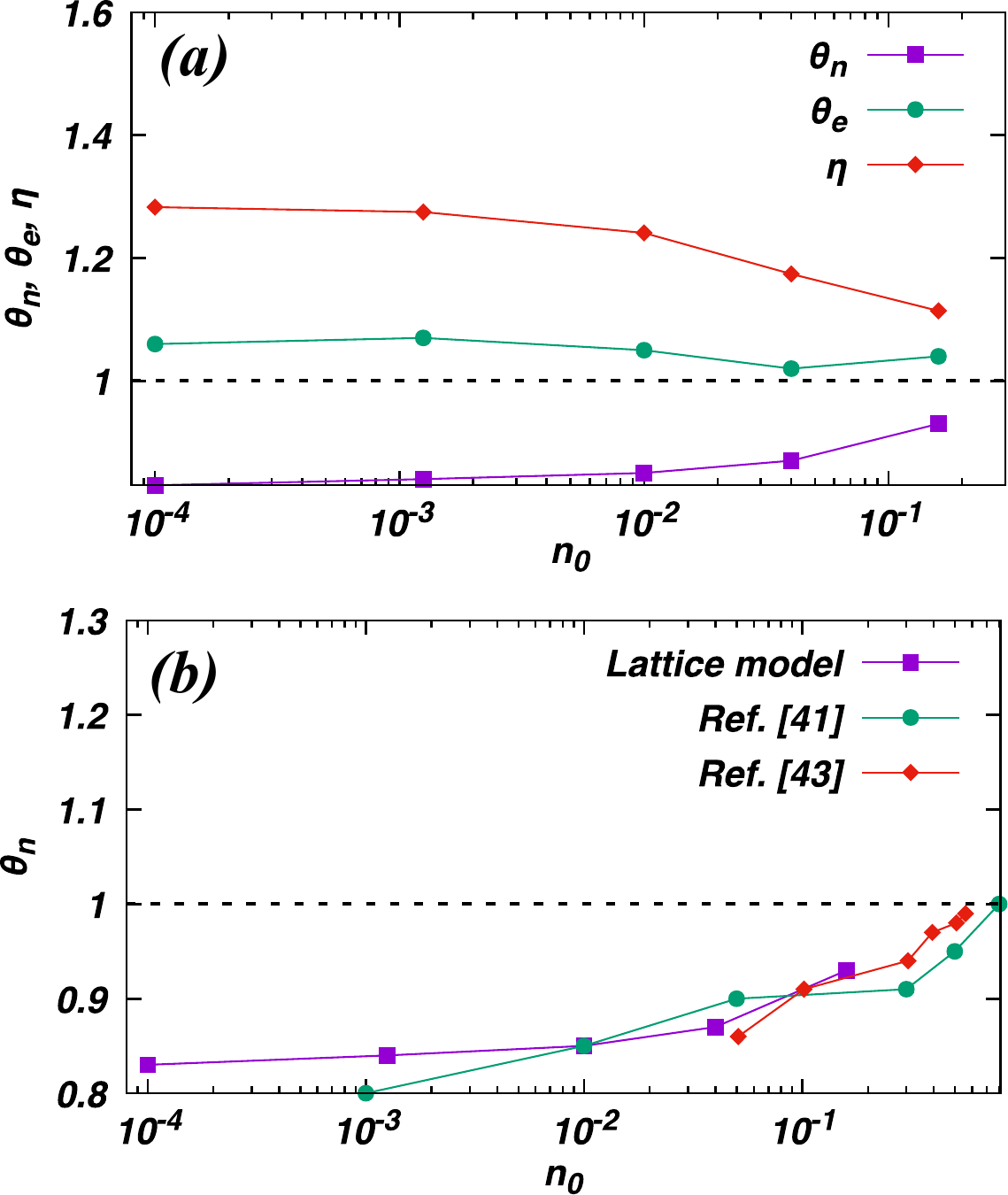}
\caption{\label{fig:mod3_exp_dendep} (a) The variation of the exponents $\theta_n$, $\theta_e$ and $\eta$ with initial density, $n_0$, for model C. The horizontal dotted line is the mean field prediction, $\theta^{\mathrm{mf}}_n=\theta^{\mathrm{mf}}_e=\eta^{\mathrm{mf}}=1$. (b) Comparison of the exponent $\theta_n$ with results of earlier simulations of BA in the  continuum~\cite{Trizac1996, Subhajit2018}.}
\end{figure}

We find that the exponents $\theta_n$, $\theta_e$, $\eta$ and $\zeta$ are density dependent~[see Table~\ref{table_mod3} and Fig.~\ref{fig:mod3_exp_dendep}(a)]. $\theta_n$ increases with the increase in density and approaches the mean field predictions $\theta^{\mathrm{mf}}_n=1$. An opposite trend is observed in the variation of exponent $\eta$ with density where it decreases with the increase in initial density but, approaches the mean field prediction $\eta^{\mathrm{mf}}=1$ with the increase in density. On the other hand, $\theta_e$ has a rather weak dependence on the initial density and is always close to the mean field result $\theta^{\mathrm{mf}}_e=1$ irrespective of the initial density. We compare our results with those for BA in the continuum~\cite{Trizac1996, Subhajit2018} in Fig.~\ref{fig:mod3_exp_dendep}(b). We find that the data are in good agreement, suggesting that the stochasticity introduced in the temporal evolution of the lattice model  is not relevant. 
\begin{table}
\caption{\label{table_mod3}Summary of the numerically obtained values of the exponents for model C.}
		\begin{ruledtabular}
		\begin{tabular}{ c c c c c }
		 $n_0$  & $\theta_n$  &  $ \eta$  &  $ \theta_e(=\eta \theta_n)$	& $\zeta$ \\
		\hline
		0.0001		&	0.83(4)		&	1.283(13)		&	1.06(6)		& -0.248(26)\\
		0.00125		&	0.84(5)		&	1.275(10)		&	1.07(7)		& -0.350(27) \\
		0.01			&	0.85(5)		&	1.241(2)		&	1.05(6)		& -0.364(6)\\
		0.04			&	0.87(6)		&	1.174(3)		&	1.02(7)		& -0.403(4) \\	
		0.16			&	0.93(5)		&	1.114(2)		&	1.04(6)		& -0.563(10) \\	
	\end{tabular}
	\end{ruledtabular}
\end{table}

\section{\label{sec:conclusion}Conclusion}

In this paper, we studied the problem of ballistic aggregation in two dimensions using three different lattice models. In all the three models, particles move, on an average, in a straight line and undergo momentum-conserving aggregation on contact. The three models differ in the shape of the particles. In Model A, the particles are point-sized and occupy a single lattice site. In model B, the shape of the aggregate is the combined shape of the two aggregating particles at the time of collision, and is a fractal. In model C, the shape of the particles are spherical, to the closest lattice approximation. For the three models, from large scale Monte Carlo simulations, we determine the exponents characterizing the power-law decay of the number density of particles, the mean energy, the fractal dimension, the correlation between the velocities of the particles constituting an aggregate and the scaling function for the mass distribution.
The results for the three models are summarized in Table~\ref{table_mod1} (model A), Table~\ref{table_mod2} and Fig.~\ref{fig:mod2_exp_dendep} (model B), Table~\ref{table_mod3} and Fig.~\ref{fig:mod3_exp_dendep} (model C).

We find that the values of the exponents are independent of the initial number density only for model A. For models B and C, the exponents are weakly dependent on the initial number density, making them non-universal. The fractal dimension in model B is, however, independent of the initial number density, within the numerical accuracy that we could achieve. In model C, the  trends in the dependence of the exponents on $n_0$  are consistent with the corresponding simulations for spherical particles in the continuum~\cite{Trizac-PRL-1995,Trizac1996,Trizac2003,Subhajit2018,rajesh-granular-prl-2014}.  While  the exponent $\theta_n$  matches closely with the continuum results [see  Fig.~\ref{fig:mod3_exp_dendep}(b)], we find that the numerical values of the exponent $\theta_e$ is less  than the continuum result~\cite{Subhajit2018} and approaches the mean field result faster. This discrepancy could be due to difficulties in measuring $\theta_e$ accurately due to strong crossovers seen in the data. We have shown that the results for the exponents in all the models, irrespective of its dependence on $n_0$,  satisfy the hyperscaling relations derived from  scaling theory. 

The fractal dimension of clusters formed by aggregation is of interest in many experiments (for example, see~\cite{sorensen2018light,zhang2020three,leggett2019motility,liu2021scale,simmler2022characterization}).
While it is to be expected that the exponents $\theta_n$ and $\theta_e$ will depend on the nature of transport and the shapes of the clusters, it is not clear whether the fractal dimension depends on transport. Fractal dimension of the cluster in two-dimensional diffusion-limited aggregation (DLA) models, where clusters grow from a nucleating center, show $d_f\simeq 1.70$~\cite{PhysRevLett.47.1400,PhysRevE.103.012138}. However, fractal dimension of clusters, when there is no nucleating center but all the aggregates undergo diffusive motion, is different from that of DLA. In the case when the diffusion constant of larger masses decreases with mass or is mass-independent,  $d_f$ has been been shown to be in the range $d_f \simeq 1.38-1.52$~\cite{PhysRevLett.51.1119,PhysRevLett.51.1123,wang2020kinetically}.
This result is close to our result for ballistic aggregation (model B) for which we found $d_f\simeq1.49$. While close, it is not clear whether the fractal dimension is different for the diffusive and ballistic models. 
The value $1.49$ is very close to that observed in sprays ($1.54$)~\cite{simmler2022characterization}, and cells ($1.5$)~\cite{liu2021scale}. It would be interesting to explore this connection further as well as understand the dependence of the fractal dimension on different mass dependent velocities, especially the limit where larger masses move faster.

The mean field approximation assumes that the velocities of the particles forming a cluster are uncorrelated. The correlations are characterized by the power-law dependence of the  speed on the mass of the aggregate: $\langle v^2(m) \rangle \sim m^{-\eta}$, with $\eta^{\mathrm{mf}}=1$. Earlier simulations of spherical particles in the continuum show that $\eta$ decreases to $\eta =\eta^{\mathrm{mf}}$ as the initial number density of particles, $n_0$, is  increased~\cite{Trizac-PRL-1995,Trizac1996,Trizac2003,Subhajit2018,rajesh-granular-prl-2014}. This lack of correlation was attributed to the increased avalanche of coagulation events that occur due to the overlap of  a newly created spherical particle with already existing particles, as the number density is increased. In this paper, we determined $\eta$ for the three models. For model A, we find that $\eta\approx 1.15$ is independent of $n_0$ and hence there is no limit in which velocities become uncorrelated. For models B and C, we find that $\eta \to \eta^{\mathrm{mf}}$ with increasing $n_0$ (see Tables~\ref{table_mod2} and \ref{table_mod3}). However, in model B there are no avalanche of collisions  while model C has avalanche of collisions. Thus, contrary to earlier conjecture, the avalanche of  collisions cannot be a necessary condition for velocities to become uncorrelated.

In contrast to BA in the continuum where the dynamics is deterministic, the temporal evolution  in the lattice models is stochastic. Each particle  moves in a straight line only on an average. In the continuum models stochasticity enters only through the  initial conditions. However, for BA in one dimension, it has been shown that the stochasticity in the dynamics not only does not affect scaling laws, the lattice models reproduce many details of the trajectory like shock positions for the same initial conditions~\cite{ostojic2004clustering,dey2011lattice}. For model C, we find that the results for $\theta_n$ match with the earlier continuum results in two dimensions for all $n_0$. We thus conclude that stochasticity in the initial conditions dominate the fluctuations induced by the dynamics. This is in sharp contrast to diffusive systems where diffusive fluctuations dominate randomness in initial conditions.

For all the three models, we measure the exponent $\zeta$ [see definition in Eq.~(\ref{eqn:small_mass_Nmt})] which characterizes the behavior of smaller mass aggregates. The exponent $\zeta$ is not easily obtained from scaling arguments and for the corresponding diffusive problem requires renormalisation group calculations~\cite{krishnamurthy2002kang,krishnamurthy2003persistence,rajesh2004survival}. 
For model A, we find that $\zeta$ is positive, implying that there is a typical time dependent mass. This is in contrast to point particles in one dimension where the mass distribution is a power law. For models B and C, we find that $\zeta$ is dependent on $n_0$. However, it is negative for all values of $n_0$, implying that the mass distribution is a power law in mass, for a given time. In addition, it would be interesting to study the
effect of spatial effects and mass-mass correlations on the exponent $\zeta$  by comparing the
results obtained from the Monte Carlo simulations with
the results from  direct numerical solution of the Smoluchowski equation, which ignores all correlations.

\begin{acknowledgments}
The simulations were carried out on the supercomputer Nandadevi at The Institute of Mathematical Sciences (IMSc). P.F would like to thank IMSc for the visiting studentship. V. V. P. acknowledges SERB Start-up research Grant No. SRG/2022/001077 for support.
\end{acknowledgments}

%

\end{document}